\documentclass[english, aip, pop, amsmath,amssymb, reprint]{revtex4-1}
\usepackage{graphicx}
\usepackage{dcolumn}
\usepackage{bm}

\newcommand{\p}{\partial}
\newcommand{\rp}{\right)}
\newcommand{\lp}{\left(}
\newcommand{\rb}{\right]}
\newcommand{\lb}{\left[}

\newcommand{\rw}{\right|}
\newcommand{\lw}{\left|}

\newcommand{\al}{\alpha}

\newcommand{\de}{\delta}
\newcommand{\ep}{\epsilon}
\newcommand{\ze}{\zeta}
\newcommand{\ta}{\theta}

\newcommand{\la}{\lambda}
\newcommand{\om}{\omega}
\newcommand{\si}{\sigma}
\newcommand{\De}{\Delta}

\newcommand{\La}{\Lambda}
\newcommand{\Om}{\Omega}

\newcommand{\colonl}{(Color online.)}
\newcommand{\Exlin}{E_x^\mathrm{lin}}
\newcommand{\dNc}{\de N_c}
\newcommand{\dNcr}{\de N_{cr}}
\newcommand{\dNel}{\de N_\mathrm{el}}
\newcommand{\dNsl}{\de N_\mathrm{sl}}
\newcommand{\erf}{\mathrm{erf}} 
\newcommand{\ftd}{p\textsc{f3d}}

\newcommand{\lde}{\lambda_{De}}
\newcommand{\loki}{\textsc{loki}}
\newcommand{\Kel}{K_\mathrm{el}}
\newcommand{\Ksl}{K_\mathrm{sl}}
\newcommand{\NDe}{N_{De}}

\newcommand{\NBel}{N_{B,\mathrm{el}}} 
\newcommand{\NBsl}{N_{B,\mathrm{sl}}} 
\newcommand{\NBx}{N_B^{\mathrm{dyn}}}

\newcommand{\ntr}{N_\mathrm{tr}}
\newcommand{\ntrc}{N_\mathrm{tr,c}}
\newcommand{\ntrel}{N_\mathrm{tr,el}}
\newcommand{\ntrsl}{N_\mathrm{tr,sl}}

\newcommand{\ntrsltwoD}{N_\mathrm{tr,sl}^\mathrm{2D}}
\newcommand{\ntrslthrD}{N_\mathrm{tr,sl}^\mathrm{3D}}

\newcommand{\nudc}{\nu_\mathrm{d,c}}

\newcommand{\nudel}{\nu_\mathrm{d,el}}
\newcommand{\nudsl}{\nu_\mathrm{d,sl}}

\newcommand{\omp}{\omega_{pe}}

\newcommand{\Rthri}{R_\mathrm{thr,i}}
\newcommand{\Rthrsl}{R_\mathrm{thr,sl}}

\newcommand{\tel}{t_\mathrm{el}}
\newcommand{\thatappr}{\hat t_\mathrm{appr}}
\newcommand{\thatex}{\hat t_\mathrm{ex}}
\newcommand{\upe}{u_\perp}
\newcommand{\vte}{v_{Te}}
\newcommand{\zeff}{Z_\mathrm{eff}}

\begin{document}

\title{Threshold for Electron Trapping Nonlinearity in Langmuir Waves} 
\author{D.\ J.\ Strozzi}
\email{strozzi2@llnl.gov}
\author{E.\ A.\ Williams}
\affiliation{Lawrence Livermore National Laboratory, Livermore, CA 94551, USA}
\author{H.\ A.\ Rose} 
\affiliation{Los Alamos National Laboratory, Los Alamos, NM 87545, USA}
\author{D.\ E.\ Hinkel} 
\author{A.\ B.\ Langdon} 
\author{J.\ W.\ Banks} 
\affiliation{Lawrence Livermore National Laboratory, Livermore, CA 94551, USA}

\pacs{52.25.Dg, 52.35.Fp, 52.35.Mw, 52.38.Bv, 52.38.-r, 52.57.-z}

\keywords{nonlinear Langmuir waves; trapped electrons; laser-plasma interaction; inertial confinement fusion; stimulated Raman scattering} 
\date{\today}

\begin{abstract}
We assess when electron trapping nonlinearity is expected to be important in
Langmuir waves. The basic criterion is that the inverse of the detrapping rate $\nu_d$ of
electrons in the trapping region of velocity space must exceed the
bounce period of deeply-trapped electrons, $\tau_B\equiv(n_e/\de n)^{1/2}2\pi/\omp$.  A unitless
figure of merit, the ``bounce number'' $N_B \equiv 1/\nu_d\tau_B$, encapsulates
this condition and defines a trapping threshold amplitude for which $N_B=1$. The
detrapping rate is found for convective loss (transverse and
longitudinal) out of a spatially finite Langmuir wave. Simulations of driven
waves with a finite transverse profile, using the 2D-2V Vlasov code \loki, show
trapping nonlinearity increases continuously with $N_B$ for transverse loss, and 
significant for $N_B \approx 1$.  The
detrapping rate due to Coulomb collisions (both electron-electron and electron-ion) is
also found, with pitch-angle scattering and parallel drag and diffusion
treated in a unified manner. A simple way to combine convective and collisional
detrapping is given. Application to underdense plasma conditions in
inertial confinement fusion targets is presented. The results show that convective transverse
loss is usually the most potent detrapping process in a single $f/8$ laser
speckle. For typical plasma and laser conditions on the inner laser cones of
the National Ignition Facility, local reflectivities $\sim3\%$ are estimated to produce significant
trapping effects.
\end{abstract}
\maketitle

\section{Introduction}

The nonlinear behavior of Langmuir waves (LWs) is a much-studied problem in
basic plasma physics from the 1950s to the present.  In this paper, we focus on
nonlinearity due to electron trapping in the LW potential well. This
intrinsically kinetic effect has motivated theoretical work such as nonlinear
equilibrium or Bernstein-Greene-Kruskal (BGK) modes \cite{bernstein-bgkmodes-pr-1957},
Landau damping reduction \cite{oneil-damping-pof-1965}, nonlinear frequency
shift
\cite{manheimer-nonlinepw-pof-1971,morales-freqshift-prl-1972,dewar-freqshift-pof-1972},
and the sideband instability
\cite{wharton-epws-pof-1968,kruer-tpi-prl-1969}. Important applications of
trapping occur in LWs driven by coherent (e.g., laser) light, including the
laser plasma accelerator \cite{tajima-laseraccel-prl-1979} and stimulated Raman
scattering (SRS) \cite{goldman-scattering-pof-1965, drake-parinst-pof-1974,
  kruer-lpi-1988}. The latter allows the prospect of laser pulse compression to
ultra-high amplitudes (the backward Raman amplifier)
\cite{malkin-ramanamp-prl-1999}. In addition, SRS is an important risk to ICF
\cite{lindl-nif-pop-2004, atzeni-icf-2004}, both due to loss of laser energy and
the production of energetic (or ``hot'') electrons that can pre-heat the fuel.
Ignition experiments at the National Ignition Facility (NIF)
\cite{moses-nif-fuscitech-2005} have shown substantial Stimulated Raman
backscatter (SRBS) from the inner cones of laser beams \cite{meezan-nif-pop-2010}. The
current study is prompted primarily by SRS-driven LW's.  Much
recent work has focused on nonlinear kinetic aspects of SRS, including
``inflation'' due to Landau damping reduction
\cite{vu-kininf-prl-2001, strozzi-srs-pop-2007, benisti-landau-prl-2009,
  benisti-srs-prl-2010, ellis-srs-pop-2012},
saturation by sideband instability \cite{brunner-valeo-prl-2004}, and LW
self-focusing in multi-D particle-in-cell simulations \cite{yin-srs-pop-2009,
  yin-srs-prl-2007, fahlen-local-pre-2011}, Vlasov simulations
\cite{banks-epws-pop-2011}, and theory \cite{rose-lwfil-pop-2008}.  One goal is
to find reduced descriptions, such as envelope equations, that approximately incorporate
kinetic effects \cite{yampolsky-epws-pop-2009, dodin-epws1-pop-2012, benisti-envelope-pop-2012}.

Our aim is to provide theoretical estimates for when electron trapping
nonlinearity is important in LW dynamics. These allow for self-consistency
checks - or invalidations - of linear calculations of LW amplitudes.  This work
is therefore not primarily intended to study nonlinear LW dynamics, although we
do present Vlasov simulations to quantify the onset of trapping in the presence
of convective transverse loss. We consider a single, quasi-monochromatic wave with
electron number density fluctuation $\de n(\vec x,t) \cos(kx-\om t)$, and
slowly-varying, unitless
amplitude $\de N \equiv \de n/n_e$ where $n_e$ is the background electron
density. We refer to an electron as ``trapped'' if it is within the phase-space
island centered about the phase velocity $v_p\equiv \om/k$ and bounded by the
separatrix in the instantaneous wave amplitude, regardless of how long it has
been there.  The dielectric response of the plasma depends on the distribution
function, and therefore manifests trapping effects only after enough time has
passed for the (typically space-averaged) distribution to be distorted. We call
such a distribution trapped or flattened, since trapping produces a plateau in
the space-averaged distribution centered at $v_p$. Deeply-trapped electrons have
an angular frequency $\om_B\equiv\omp \de N^{1/2}$ ($\omp^2=n_ee^2/\ep_0m_e$
defines the plasma frequency in SI units), known as the bounce frequency,
corresponding to a bounce period $\tau_B\equiv 2\pi/\om_B \propto \de N^{-1/2}$.
In our language, an electron is trapped instantaneously, but a distribution
becomes trapped over a time $\sim\tau_B$. For a process that detraps electrons
at a rate $\nu_d$, the unitless ``bounce number'' $N_B \equiv 1/\nu_d\tau_B$
measures how many bounce orbits a trapped electron completes before being
detrapped.

Our estimates stem from the assumption that nonlinear trapping
effects are significant when $N_B$ is roughly unity. Trapping nonlinearity develops
continuously with wave amplitude, and is not an instability with a hard
threshold.  Vlasov simulations presented in Sec.\ \ref{s:loki} of driven LWs with a finite
transverse profile demonstrate this.  In addition, transit-time
damping calculations \cite{rose-ttd-dpp-2006} show the reduction in
Landau damping varies continuously with $N_B$ and obtains a 2x reduction for
$N_B \approx 1$.  Bounce number estimates are qualitative and
demonstrate basic parameter scalings.  The quantitative role of trapping depends
on the specific application.

We consider two detrapping processes: convective loss and Coulomb
collisions.  For a LW of finite spatial extent, electrons enter and leave the
wave from the surrounding plasma (assumed here to be in thermal equilibrium,
i.e.\ Maxwellian).  Trapping will only be effective if these electrons complete
a bounce orbit before transiting the wave.  We find the detrapping rate for both
longitudinal end loss, which can be important in finite-domain 1D kinetic
simulations, and for transverse side loss in 2D and 3D.  To quantify the effect
of trapping in a LW with finite transverse extent, we perform 2D-2V simulations
with the parallel Vlasov code \loki \cite{banks-epws-pop-2011, banks-loki-ieeetps-2010} of
a LW driven by an external field with a smooth transverse profile. Our
results are in qualitative agreement with Sec.\ IV of Ref.\
\onlinecite{banks-epws-pop-2011}. That work considered a free LW excited by a driver
of finite duration, while we consider a driver that remains on.

We present a unified calculation of collisional detrapping due to electron-ion
and electron-electron collisions, including both pitch-angle scattering and
parallel slowing down and diffusion. This relies on the fact that (see the
Appendix) the distribution in the trapping region can be Fourier decomposed into
modes $\sin[n\pi((v_x-v_p)/v_{tr}+1/2)]$ for $n=1,3,...$, and the diffusion rate of mode $n$ is
proportional to $n^2$.  After a short time, only electrons in the fundamental
$n=1$ mode remain trapped. The
collisional detrapping rate scales as $1/\de N$, since the trapping width in
velocity increases with wave amplitude. We discuss two ways to compare the relative
importance of detrapping by side loss and collisions, which is complicated by
their different scaling with $\de N$. 

Our calculations are applied to ICF plasma conditions, particularly LW's driven
by stimulated Raman backscatter (SRBS) on the NIF. Transverse side loss out of
laser speckles in a phase-plate-smoothed beam is generally a more effective
detrapping process than collisions.  The threshold $\de N$ for trapping to
overcome side loss decreases with density and increases with temperature, while
the collisional threshold decreases with density and slightly increases with
temperature. For conditions typical of backscatter on NIF ignition experiments,
namely $T_e$=2 keV and $n_e=0.1n_{cr}$ with $n_{cr}\equiv \om_0^2\ep_0m_e/e^2$ the critical density
for laser light of wavelength 351 nm, a reflectivity of $(5\times10^{13}$ W
cm$^{-2}/I_0)^2$ produces linear Langmuir waves above the side loss
threshold. Such values are likely to occur in intense speckles. We also show that smoothing by spectral
dispersion (SSD) \cite{skupsky-ssd-jap-1989} is ineffective at detrapping in
NIF-relevant conditions.

The paper is organized as follows. Section \ref{s:gen} provides some general
considerations on our detrapping analysis. We present in Sec.\ \ref{s:convloss} convective
loss calculations for both longitudinal (end) and transverse (side)
loss. Section \ref{s:loki} contains Vlasov simulations with the \loki\ code which study the
competition of trapping and side loss. Detrapping by Coulomb collisions is
treated in Sec.\ \ref{s:coll}. Our results are applied to SRBS in underdense ICF
conditions in Sec.\ \ref{s:icf}. We conclude in Sec.\ \ref{s:conc}. The Appendix presents
details of our collisional derivation and discusses the validity of our
Fokker-Planck model.

\section{General Considerations} \label{s:gen} 

This section presents our overall framework for estimating the trapping
threshold, and lays out some definitions. Consider the trapped electrons in a LW field, attempting to undergo
bounce orbits. There is a time-dependent condition for trapping to distort the distribution
significantly, even in the absence of any detrapping process.  For instance, if a
LW is suddenly excited in a Maxwellian plasma, electrons execute bounce orbits
according to what we call the dynamic bounce number
\begin{equation}
  \label{eq:NBx}
  \NBx(t) = \int_0^t {dt' \over \tau_B(t')}.
\end{equation}
The time dependence of $\tau_B$ allows for a slowly-varying wave amplitude $\de
n(t)$. Vlasov simulations presented in Sec.\
\ref{s:loki} show that trapping
starts to significantly affect the dielectric response when $\NBx \approx 0.5$. That
is, it takes a finite time for the distribution to reflect
trapping. The early works of Morales and O'Neil \cite{oneil-damping-pof-1965,
 morales-freqshift-prl-1972} indicate such behavior, where the damping
rate and frequency shift evolve over several bounce periods until approaching
steady values as the system reaches a Bernstein-Greene-Kruskal (BGK) state
\cite{bernstein-bgkmodes-pr-1957}.

To estimate the threshold for trapping to overcome a detrapping process, we
assume the wave has been present long enough that $\NBx\gtrsim1$.  The
distribution has had enough time to become flattened, to the extent the
detrapping process allows. For flattening to occur,
an appreciable fraction of trapped electrons must remain so for about a bounce
period before being detrapped.  We are interested in the number of electrons in
the trapping region, and how long they stay there.
 
We define the ``trapping region'' to extend from $u=u_p\pm u_{tr}/2$ where
$u_{tr}\equiv 4(k\lde)^{-1}\de N^{1/2}$ is the full width of the phase-space
trapping island and $\lde\equiv \vte/\omp$ with
$\vte\equiv(T_e/m_e)^{1/2}$. Throughout this paper, we use 
\begin{equation}
u_X \equiv v_X/\vte
\end{equation}
to denote the scaled velocity $v_X$ for various subscripts $X$. Let $\ntr(t)$ denote the fraction of electrons in the trapping region at the initial time $t=0$, that continuously remain so to some later time $t$ (note $\ntr(t=0)=1$). At $t=0$ we take the electron distribution to be Maxwellian. The fact that only some electrons in the trapping region lie within the separatrix (depending on their initial phase $kx$) is not relevant, since all the detrapping processes considered here are insensitive to the electron's phase in the wave.  That is, the rate at which electrons leave the trapping region is independent of $kx$.

The detrapping rate $\nu_d$ is defined by assuming exponential decay for the trapped fraction: $\ntr=e^{-\nu_dt}$. We allow for several
independent detrapping processes to occur simultaneously, in that the overall
detrapping rate $\nu_{d,O}$ is the sum of the rates $\nu_{d,i}$ for
each $i$th process considered separately. Since a detrapping process generally
does not strictly follow exponential decay, we choose a critical fraction
$\ntr^*$, which obtains for a critical time $t=t^*$, and let
$\nu_d=\ln(1/\ntr^*)/t^*$. $\nu_d$ is independent of $\ntr^*$ for
exponential decay. We set $\ntr^*=1/2$ in what follows. Given the approximate nature of our calculation, further refinement of $\nu_d$ has little value.

In the literature, detrapping processes are sometimes approximated by a 1D kinetic equation with a Bhatnagar-Gross-Krook
relaxation (or simply a Krook) operator \cite{bgk-op-pr-1954}:
\begin{equation}
  \label{eq:bgk}
  \lb \p_t + v\p_x - (e/m_e)E\p_v \rb f=\nu_K\cdot(nf_0/n_0-f).
\end{equation}
The linear electron susceptibility $\chi$ for this kinetic equation is
\begin{equation}
  \label{eq:chi}
  \chi(\om,k) = -{Z'(\ze) \over 2(k\lde)^2} \lb 1+i {\nu_K \over k\vte\sqrt2}Z(\ze) \rb ^{-1},
\end{equation}
where $\ze\equiv\om/k\vte\sqrt2$ and $Z$ is the plasma dispersion function
\cite{fried-zfunc-1961}.  The Krook operator relaxes the electron distribution function $f$ to an
equilibrium $f_0$, and locally conserves number density $n=\int dv\,f$. The
above operator does not conserve momentum or energy, although it can easily be
generalized to do so. In a 1D-1V system, a Krook operator can mimic detrapping by
transverse convective loss (a higher space-dimension effect) or Coulomb
collisions (a higher velocity-dimension effect), such as in Ref.\
\onlinecite{rose-nonlinEPW-pop-2001}. Any perturbation from $f_0$ decays
exponentially at the rate $\nu_K$, so $\nu_d=\nu_K$ for such an operator.  This
is especially useful for a detrapping process which has $\nu_d$ independent of
wave amplitude; this is the case for convective loss but not for collisions (as
shown below). SRS simulations with a 1D Vlasov code and Krook operator, and its
suppression of kinetic inflation, are presented in Ref.\
\onlinecite{strozzi-vlasov-shoucri-book}. In this paper, we do not use a Krook
operator to model detrapping, although we do use one in our 2D Vlasov
simulations to make them effectively finite in the transverse direction (a
purely numerical purpose), and to include collisional LW damping in our
application to ICF conditions in Sec.\ \ref{s:icf}.

We take the bounce period of all trapped electrons to be $\tau_B$, the result
for deeply-trapped electrons. The actual period slowly increases to infinity for electrons near the separatrix. We then define the bounce number for process $i$ as 
\begin{equation}
  \label{eq:2}
 N_{B,i} \equiv { 1 \over \nu_{d,i}\tau_B} = \lb {\de N \over \de N_i} \rb ^{p_i}.
\end{equation}
We have expressed $N_{B,i}$ as a ratio of the LW amplitude to a ``threshold''
amplitude $\de N_i$, to some power $p_i$. Recall that trapping effects like the
Landau damping reduction develop continuously with $\de N$, so the threshold for
trapping nonlinearity is not a hard one. Besides the $\de N^{-1/2}$ dependence
of $\tau_B$, $\nu_{d,i}$ also depends on $\de N$ in a process-dependent way. For
$\nu_{d,i}$ independent of wave amplitude, which we show below is the case for
convective loss, the power $p_i=1/2$. This is not the case for detrapping by Coulomb
collisions, which is shown in Sec.\ \ref{s:coll} to have $p_i=3/2$. The overall detrapping rate
$\nu_{d,O}=\sum_i\nu_{d,i}$, gives an overall bounce number via
$N_{B,O}^{-1}=\sum_iN_{B,i}^{-1}$. We also define an overall threshold amplitude
$\de N_O$ such that $N_{B,O}[\de N=\de N_O]=1$; it is \textit{not} generally
true that $\de N_O=\sum_i\de N_i$.

\section{Convective Loss: Theory} \label{s:convloss} 

In a LW of finite spatial extent, electrons remain in the trapping region only
until they transit the wave. This detrapping manifests itself by longitudinal
loss out of the ends of the wavepacket (the $x$ direction for our field
representation $\cos(kx-\om t)$), as well as transverse loss out the sides. End
loss is found by considering a wavepacket of length $L_{||}$ and infinite
transverse extent.  We work in the rest frame of the wavepacket, which may
differ from the lab frame depending on application.  For instance, a free LW
propagates at group velocity $v_g = 3\vte^2/v_p$ for $k\lde\ll1$, while a LW
driven by a driver fixed in the lab frame (such as the ponderomotive drive in
SRS) will essentially be at rest. For $v_{tr}\ll v_p$ we can treat all trapped
electrons as moving forward at $v_p$.  Thus for end loss
$\ntrel=1-v_pt/L_{||}$. To find $\nudel$, we take $\ntrel^*=1/2$, which gives
$\tel^*=L_{||}/v_p$ and $\nudel=\Kel v_p/L_{||}$ with $\Kel=\ln 2$. The bounce
number for end loss is $\NBel=[\de N/\dNel]^{1/2}$, with exponent $p_{\mathrm{el}}=1/2$
and threshold amplitude $\dNel=[2\pi \Kel u_p\lde/L_{||}]^2$. In practical units
$\dNel=(1.05\times 10^{18}/n_{e,cc} )T_{e,kV}(u_p/ L_{|| ,\mu m})^2$ where
$n_{e,cc}$ is in cm$^{-3}$, $T_{e,kV}$ is in keV, and $L_{||,\mu m}$ is in
$\mu$m.

For transverse side loss, consider a cylindrical wavepacket of transverse
diameter $L_\perp$ and infinite longitudinal length. In $N$ total spatial
dimensions, the cylinder has an $N-1$ dimensional cross-section. Electrons with
a Maxwellian distribution are transiting the cylinder, with unnormalized
distribution $f_\perp=\upe ^{N-2}\exp[-\upe ^2/2]$ where $\upe $ is the
transverse speed and $f_\perp d\upe $ is the number of electrons per $d\upe
$. The average $\upe  = ([\pi/2]^{1/2}, [8/\pi]^{1/2})=(1.25,1.60)$ for
$N=(2,3)$, indicating that detrapping is faster in 3D than in 2D.  

\begin{figure}
  \centering
  \includegraphics[width=3.3in]{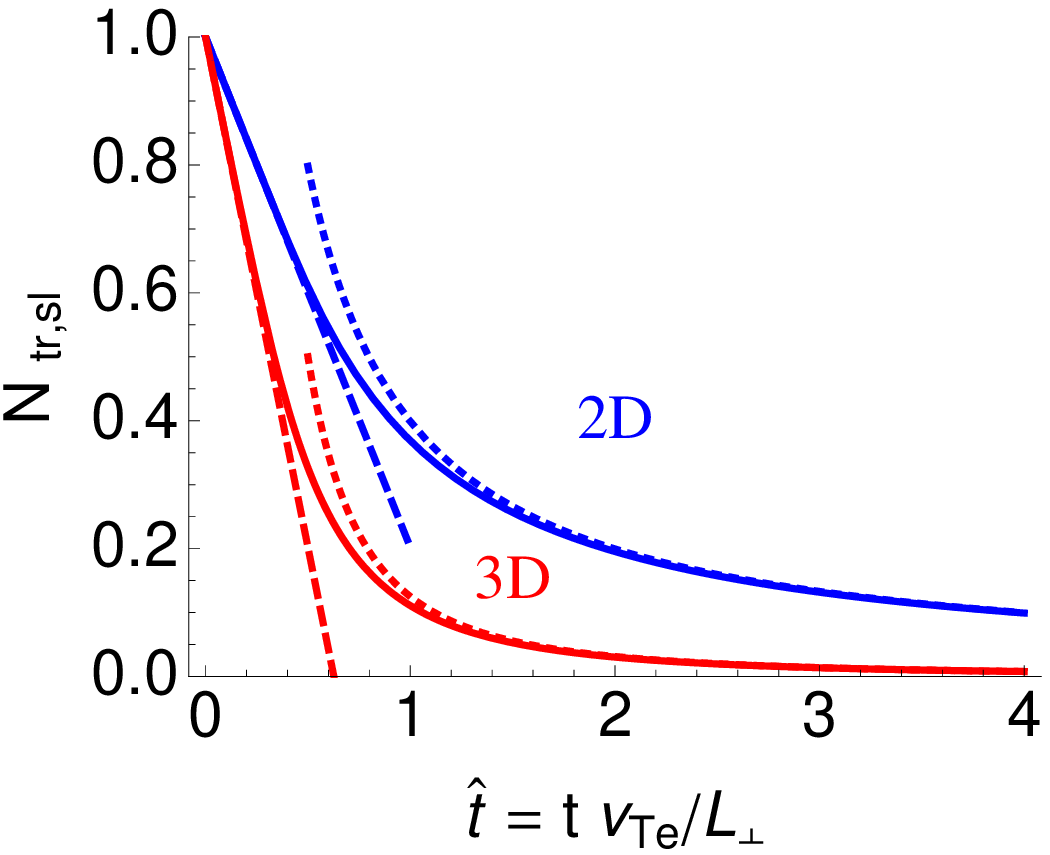}
  \caption{\colonl\ Trapped electron fraction $\ntrsl$ due to transverse side
    loss for a 2D plane (blue) and 3D cylinder (red) region. The dashed curves
    are the appropriate early and late time limits. See Eqs. (\ref{eq:ntrsl2})
    through (\ref{eq:ntrslapp3dlate}). }
  \label{fig:ntrsl}
\end{figure}

We find the number of initially trapped electrons $(|\vec x_\perp|<L_\perp/2)$, that remain so after time $t$,
by summing the fraction of electrons with a given $u_\perp$ that remain trapped,
times $f_\perp$. All electrons with $|u_\perp|>1/\hat t$ with $\hat
t=t\vte/L_\perp$ have escaped, so this sets
the limits of integration. In 2D, the trapped fraction is $(1-|u_\perp|\hat t)$ for
$|u_\perp|<1/\hat t$, and
the total trapped fraction is
\begin{eqnarray} \label{eq:ntrsl2}
  \ntrsltwoD &=& (2\pi)^{-1/2} \int_{-1/\hat t}^{1/\hat t} du_\perp\ e^{-u_\perp^2/2} \lb
  1-|u_\perp|\hat t \rb \\
 &=& \erf \lb 1/\hat t\sqrt 2 \rb + (2/\pi)^{1/2}\hat t(e^{-1/2\hat t^2}-1).
\end{eqnarray}
In 3D we obtain
\begin{eqnarray}  \label{eq:ntrsl3}
  \ntrslthrD &=& \int_{0}^{1/\hat t} du_\perp\ u_\perp\ e^{-u_\perp^2/2} \cdot \nonumber \\
  &&\lb 1-{2\over\pi}(\arcsin[u_\perp\hat t]+u_\perp\hat t[1-(u_\perp\hat t)^2]^{1/2}) \rb .
\end{eqnarray}
The factor in square brackets is the trapped fraction. The limiting forms are
\begin{eqnarray}
  \ntrsltwoD(\hat t \ll 1) &\approx& 1-[2/\pi]^{1/2} \hat t, \\
  \ntrsltwoD(\hat t \gg 1) &\approx& 1/[2\pi]^{1/2} \hat t, \\
  \ntrslthrD(\hat t \ll 1) &\approx& 1-[8/\pi]^{1/2} \hat t, \\
  \ntrslthrD(\hat t \gg 1) &\approx& 1/8 \hat t^2. \label{eq:ntrslapp3dlate}
\end{eqnarray}
In both limits the decrease is more rapid in 3D than in 2D. Figure \ref{fig:ntrsl} displays the various formulas for $\ntrsl(t)$.

The resulting detrapping rate, based on $\ntrsl=1/2$, is 
\begin{equation}
  \label{eq:nudsl}
 \nudsl = {\Ksl \vte \over L_\perp}
\end{equation}
with $\Ksl =(1.02,2.08)$ in (2D, 3D).  As expected, the 3D detrapping
rate is faster. The 3D detrapping rate exceeds the 2D one by a larger factor
than the average transverse speed because the faster electrons leave first, and
the relative surplus of electrons in 3D over 2D (proportional to $\upe $)
increases with transverse speed. A wavepacket with asymmetric (e.g.\ elliptical)
cross-section should have a rate between the 2D and 3D result with $L_\perp$
taken as the shortest transverse length. In a laser beam smoothed with phase
plates, elliptical speckles can be produced by certain polarization-smoothing
schemes or a non-spherical lens; Langmuir waves driven by SRS in such speckles
would also acquire an elliptical cross-section.

Comparing the end loss and side loss rates gives
\begin{equation}
  {\nudel \over \nudsl} = {\Kel \over \Ksl} {v_p\over \vte}{L_\perp\over L_{||}}.
\end{equation}
$v_p$ is in the wavepacket frame. For the LW to not experience strong Landau damping, we have $v_p>\vte$. $L_\perp/L_{||}$ depends on the physical situation (laser speckles are
discussed in Sec.\ \ref{s:icf}).  The bounce number for side loss is analogous
to end loss: $\NBsl=[\de N/\dNsl]^{1/2}$, with exponent $p_{\mathrm{sl}}=1/2$ and
threshold amplitude $\dNsl=[2\pi \Ksl \lde/L_\perp]^2$. In practical units and
for the 3D $\Ksl$,
$\dNsl=(9.44\times 10^{18}/n_{e,cc} )T_{e,kV}/ L^2_{\perp ,\mu m}$.

\section{Vlasov simulations of convective side loss} \label{s:loki} 

In this section, we quantify the competition between convective side loss and
electron trapping in a driven Langmuir wave.  We use the parallel, 2D-2V Eulerian Vlasov code
\loki \cite{banks-loki-ieeetps-2010}. This code employs a finite-volume method
which discretely conserves particle number. The discretization uses a 
fourth-order accurate approximation for well-resolved features, and smoothly
transitions to a third-order upwind method as the size of solution features 
approaches the grid scale. This construction enables accurate long-time integration
by minimizing numerical dissipation, while retaining robustness for nonlinearly
generated high frequencies. As a result, the method is not strictly monotone- or 
positivity-preserving, nor does it eliminate the so-called recurrence problem. 
This occurs at a recurrence time of $t_{rec}=\la/\De v$ when 
further linear evolution of a sinusoidal perturbation cannot be represented
on a given grid.

Our simulations are 1D or 2D, with $x$ the longitudinal coordinate as above, and
$y$ the transverse coordinate. Only electrons are mobile, there is a
fixed, uniform neutralizing background charge, and there is no
magnetic field. The total electric field is $\vec E = E_x\hat
x+E_y\hat y = \vec E_d+\vec E_i$, where the internal electric field $\vec E_i=-\nabla\phi_i$
and $\nabla^2\phi_i=-\rho/\ep_0$. The external driver field is $\vec E_d=E_d\hat x$ with
\begin{equation}
  \label{eq:1}
  E_d = E_0 A(t)h(y)\cos(k_0x-\om_0t).
\end{equation}
There is no $y$ component to the driver field, which would be needed if the driver
were derived from a scalar potential. The temporal envelope $A(t)$ ramps up from zero to unity over a time $50/\omp$
and then stays constant. The transverse profile $h(y)$ is
\begin{eqnarray}
  \label{eq:5}
  h(y) &=& \cos^2{2\pi y\over L_y} = \tfrac{1}{2}\lp 1+\cos k_1y \rp, \quad |y| <
  {L_y\over4} \\
  && 0 \quad \mathrm{otherwise}.
\end{eqnarray}
$k_1\equiv 4\pi/L_y$. 

\begin{figure*}
  \centering
  \includegraphics[width=7in]{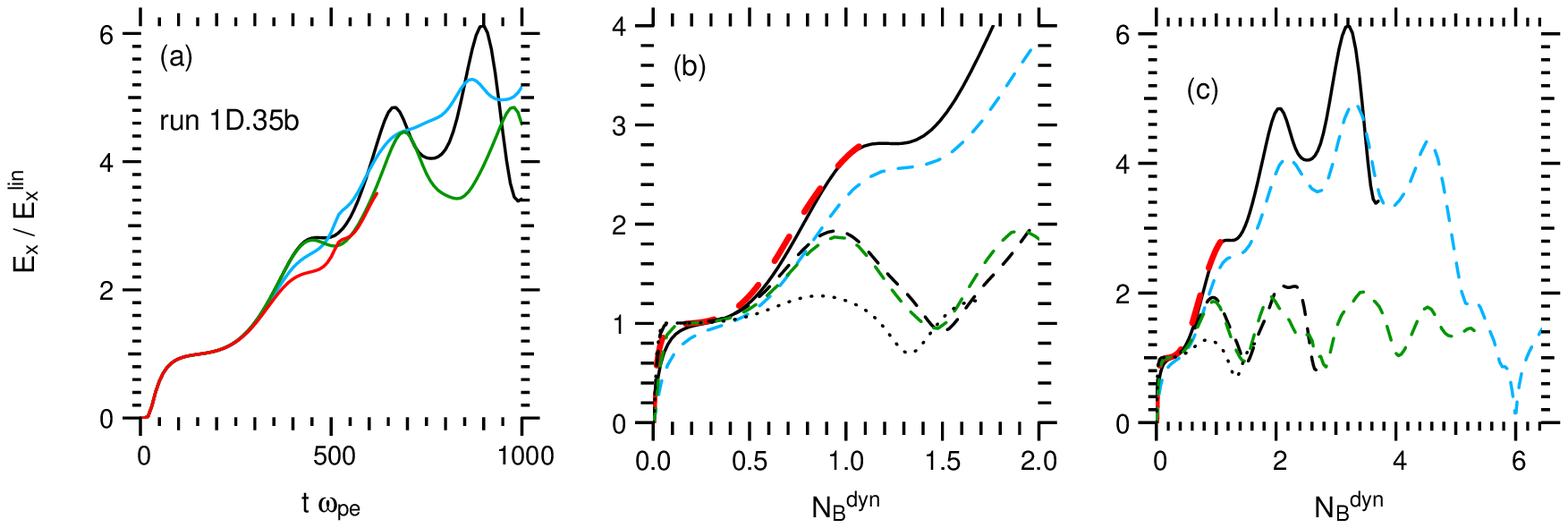}
  \caption{\colonl\ (a) Amplitude of $k=k_0$ mode of $E_x$, scaled to linear response, vs.\
   time for 1D \loki\ case \texttt{1D.35b} and four different resolutions:
   $(N_x,N_{vx})=(32, 768)$ (black), $(64, 768)$ (green), $(32, 384)$ (blue),
   and $(64, 384)$ (red). (b) $E_x$ vs.\ dynamic bounce number
   $\NBx$ from Eq.\ (\ref{eq:NBx}) using $E_x(t)$ from 1D \loki\ runs in Table \ref{tab:1D} (see table
   for curve meanings). (c) panel (b) for expanded domain. The black curve is the only run in (a) that appears in
   (b) and (c).}
  \label{fig:exlin_t}
\end{figure*}

The numerical aspects of our runs are as follows.  The $x$ domain extends for one driver wavelength,
with periodic boundaries for fields and particles. $N_x=32$ zones
in $x$ was used for all runs in this paper, except for two $N_x=64$ cases in
Fig.\ \ref{fig:exlin_t}(a). 2D runs had periodic boundaries for fields and particles at
$|y|=L_y/2$. A Krook operator with $\nu_K(y)=0$ for $|y|<0.4L_y$ and rising
rapidly in the boundary region $0.4<|y|/L_y<0.5$ was used to relax the
distribution to the initial Maxwellian near the transverse boundaries. The runs
were thus effectively finite in $y$. We used $N_y=$ 11 to 45 zones in $y$, with
more used for larger $L_y$ and to check convergence. The $v_x$ and $v_y$ grids
both extended to $\pm 7v_{Te}$. $N_{vy}=32$ zones in $v_y$ were used
throughout. $N_{vx}$ is set by two requirements: the trapping region must be
adequately resolved, and recurrence phenomena must not be significant. We found
$\De v_x \sim0.1 v_{tr}$ was sufficient to give converged results.  \loki's
advection scheme is designed to mitigate aliasing problems, and we only saw
modest effects related to it when comparing runs with different $N_{vx}$. The convergence of our numerical
results is shown in Fig.\ \ref{fig:exlin_t}(a). The black curve is typical: it
uses $N_x=32$ and has a typical
$\De v_x / v_{tr}$, which we kept similar by varying $N_{vx}$ with wave amplitude and $k_0$.

We first present 1D runs with $E_y=0$ and $h(y)=1$, which are detailed in Table
\ref{tab:1D}. From linear theory with $A(t)=1$, $E_x=\Exlin\cos(k_0x-\om_0t)$ where
\begin{equation}
  \label{eq:ET}
  {\Exlin \over E_0} = \lw {1\over 1+\chi} \rw = \lb (1+\mathrm{Re}\chi)^2 +
  (\mathrm{Im}\chi)^2 \rb^{-1/2}.
\end{equation}
$\chi$ is the linear electron susceptibility from Eq.\ (\ref{eq:chi}) with
$\nu_K=0$, evaluated at the driver $k_0$ and $\om_0$. We chose
$\om_0$ to give nearly the maximum $\Exlin$ for a given $k_0$. For $k_0\lde<0.53$,
a linearly resonant $\om_0$ exists where $1+\mathrm{Re}\chi=0$; the maximum $\Exlin$ then occurs
close to this point. No linear resonance exists for $k_0\lde>0.53$, which is called the
loss of resonance \cite{rose-nonlinEPW-pop-2001}. Some $\om_0$ still maximizes $\Exlin$ in this regime. The non-resonant case differs from the
resonant one, in that reducing $\mathrm{Im}\chi$ and Landau damping, e.g.\ by flattening the
distribution at the phase velocity by electron trapping or some other means,
does \textit{not} lead to a large enhancement in the Langmuir wave response to
an external drive. The term $1+\mathrm{Re}\chi$ in Eq.\ (\ref{eq:ET}) keeps
$\Exlin$ finite even if $\mathrm{Im}\chi=0$. For the parameters of the run
\texttt{1D.7a}, we find $\Exlin/E_0=1.80$ for the full, complex $\chi$, while
setting $\mathrm{Im}\chi=0$ slightly increases it to $\Exlin/E_0=2.01$. 

Similar logic applies to kinetic inflation of stimulated Raman
scattering. Electron trapping and the resultant Landau damping reduction can
greatly increase the scattering at a resonant wavelength. However, scattering at
a non-resonant wavelength is not subject to inflation, and can even decrease, due to reducing
$\mathrm{Im}\chi$. Non-resonant SRS can occur in a situation seeded away from
resonance \cite{ellis-srs-pop-2012}, or
if the plasma conditions are such that no resonance exists for any scattered
wavelength, namely high $T_e$ and low $n_e$.

Figure \ref{fig:exlin_t} presents the results of our 1D runs. Panel (a) shows
the time evolution of the amplitude of $E_x$ for $k=k_0$, normalized to the
linear value from Eq.\ (\ref{eq:ET}). Early in time ($\omp t=100-200$) the
linear response is achieved, which validates the linear dispersion and
properties of \loki\ when using the chosen grid resolution. As time progresses
the response increases due to the damping reduction, and then oscillates due to
the interplay of the frequency shift and the fixed driver. Similar behavior was
seen in Ref.\ \onlinecite{yampolsky-epws-pop-2009}.  We plot the results vs.\
the dynamic bounce number $\NBx$ from Eq.\ (\ref{eq:NBx}), using the
time-dependent $E_x$, in the center and right panels. $\NBx$ is thus a
trapping-based re-scaling of time. The other runs from Table \ref{tab:1D} are
included as well. The driver strength $E_0$ was chosen in runs \texttt{1D.35b},
\texttt{1D.5a}, and \texttt{1D.7a} to give similar bounce periods. In all cases,
the linear response is achieved after a transient period related to driver
turn-on, until $\NBx\approx0.5$. After this point the response increases, until
the frequency shift develops at $\NBx\approx1$. As $k_0\lde$ increases, the
enhancement above linear response decreases. This is likely due to the rapid
increase of the frequency shift with $k\lde$, as shown by most theoretical
calculations, e.g.\ Ref.\ \onlinecite{morales-freqshift-prl-1972}. For
$k_0\lde=0.7$, there is a slight enhancement to 1.3x the linear response,
followed by a dip to about 0.7x and subsequent oscillation about unity. This
lack of significant trapping nonlinearity agrees with the above discussion of
the non-resonant regime.

\begin{table}
\caption{\label{tab:1D}1D \loki\ runs with no transverse driver profile $h(y)$. $\tilde
  E_0= E_0 e / m_ev_{Te}\omp$. $\tau_B^\mathrm{lin}$ is found using $\Exlin$.}
\begin{ruledtabular}
\begin{tabular}{ccccccc}
Run & $k_0\lde$ & $\om_0/\omp$ & $\Exlin/E_0$ & $\tilde E_0$
& $\tau_B^\mathrm{lin}\omp$ & plot curve \\
\hline
\texttt{1D.35a} & 0.35 & 1.22 & 11.9 & 1.25$\times10^{-5}$ & 871 & red dash \\  
\texttt{1D.35b} & ''   & ''   & ''   & 5$\times10^{-5}$    & 436 & solid black \\ 
\texttt{1D.35c} & ''   & ''   & ''   & 2$\times10^{-4}$    & 218 & blue dash \\ 
\texttt{1D.5a}  & 0.5  & 1.44 & 3.22 & 1.3$\times10^{-4}$  & 434 & black dash \\ 
\texttt{1D.5b}  & ''   & ''   & ''   & 5.2$\times10^{-4}$  & 217 & green dash \\ 
\texttt{1D.7a}  & 0.7  & 1.79 & 1.80 & 1.7$\times10^{-4}$  & 430 & black dot   
\end{tabular}
\end{ruledtabular}
\end{table}

\begin{table}
\caption{\label{tab:2D}2D \loki\ runs with transverse driver profile $h(y)$. All runs
  have $k_0\lde=0.35$, $\om_0/\omp=1.22$, and  $E_0 e /
  m_ev_{Te}\omp=2\times10^{-4}$, the same as run \texttt{1D.35c}.}
\begin{ruledtabular}
\begin{tabular}{cccc}
Run & $L_y/\lde$ & $\NBsl=L_y/884\lde$ & plot curve \\
\hline
\texttt{2D100}  & 100  &  0.113 & red \\  
\texttt{2D200}  & 200  &  0.226 & dark blue \\ 
\texttt{2D400}  & 400  &  0.452 & green \\ 
\texttt{2D800}  & 800  &  0.905 & magenta \\ 
\texttt{2D1200} & 1200 &  1.357 & blue  
\end{tabular}
\end{ruledtabular}
\end{table}
 
From Eq.\ (\ref{eq:nudsl}), the 2D side loss
rate is $\nudsl=4.08\vte/L_y$, where we have taken $L_\perp=L_y/4$, the full-width at
half-max of $h(y)$. The side loss bounce number is then
\begin{equation}
  \label{eq:foo}
  \NBsl = {L_y \over 25.6\lde} \de N ^{1/2}.
\end{equation}
Recall that electrons feel the total electric field $\vec E$ (drive plus
interal), and $\de N$ is an equivalent density fluctuation.  Gauss's law gives $\de N =
k_0\lde\cdot\tilde E_x^0$, where $E_x^0$ is the amplitude of the $k_0$ Fourier
mode of the on-axis field $E_x(y=0)$, and $\tilde E = Ee / m_e\vte\omp$ denotes a normalized
field. Using the linear response from Eq.\ (\ref{eq:ET}), we obtain the linear
estimate
\begin{equation}
  \label{eq:NBsllin}
  \NBsl = {L_y \over 25.6\lde} \lw {k_0\lde \over 1+\chi} \rw^{1/2} \tilde E_0 ^{1/2}.
\end{equation}
The 2D \loki\ runs are listed in Table \ref{tab:2D}. All runs used
$k_0\lde=0.35$, $\om/\omp=1.22$, and  $\tilde E_0 =2\times10^{-4}$, the same as
run \texttt{1D.35c}. For these values, our linear estimate becomes $\NBsl = L_y
/ 884\lde$.

\begin{figure}
  \centering
  \includegraphics[width=3.3in]{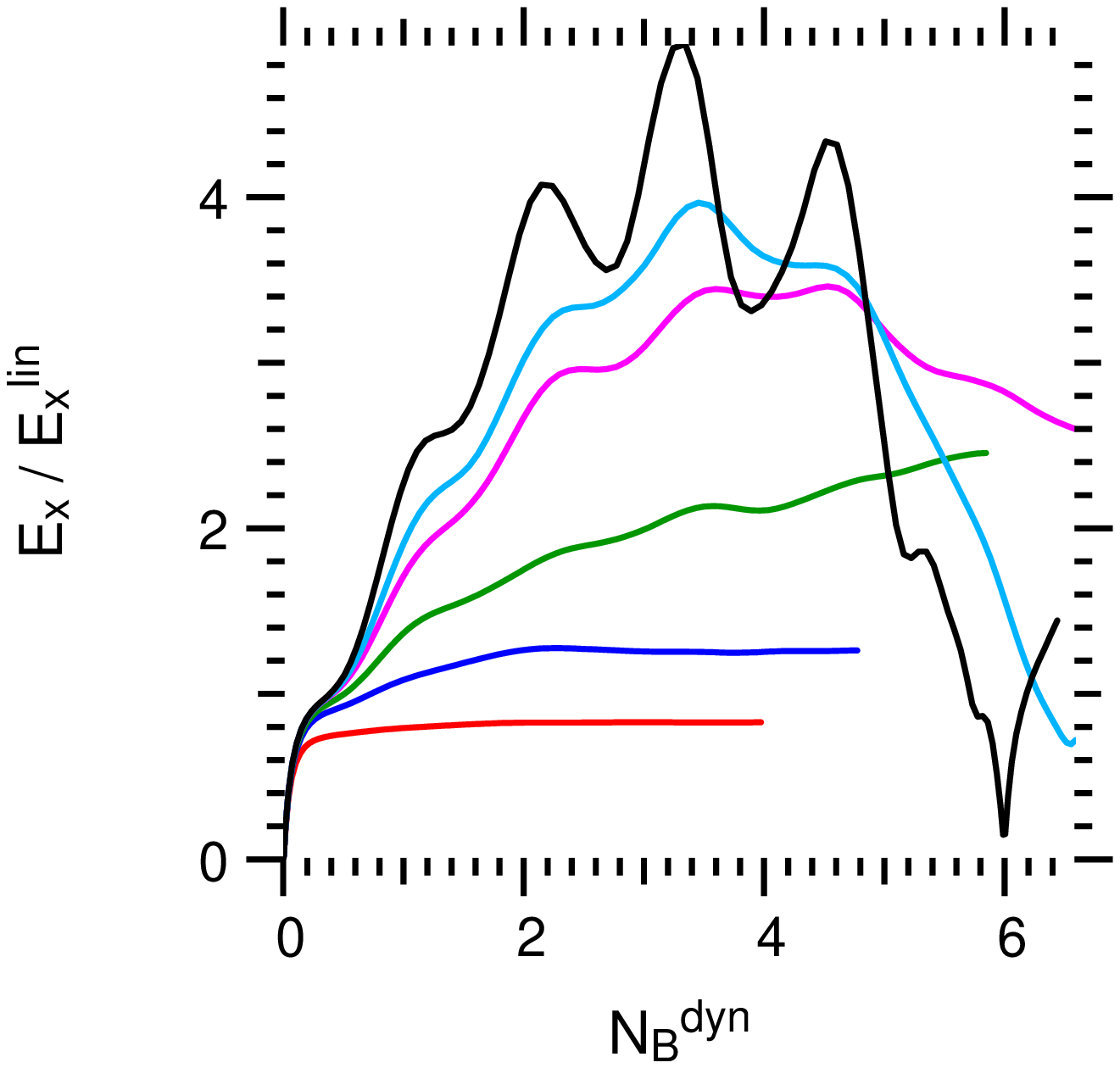}
  \caption{\colonl\ Amplitude of $k=k_0$ mode of $E_x(y=0)$ for 2D \loki\ runs with transverse driver profiles
    $h(y)$ with various $L_y$. Run parameters and curve meanings are given in
    Table \ref{tab:2D}. Black curve is 1D run \texttt{1D.35c}.}
  \label{fig:exlin_Ly}
\end{figure}

\begin{figure}
  \centering
  \includegraphics[width=3.3in]{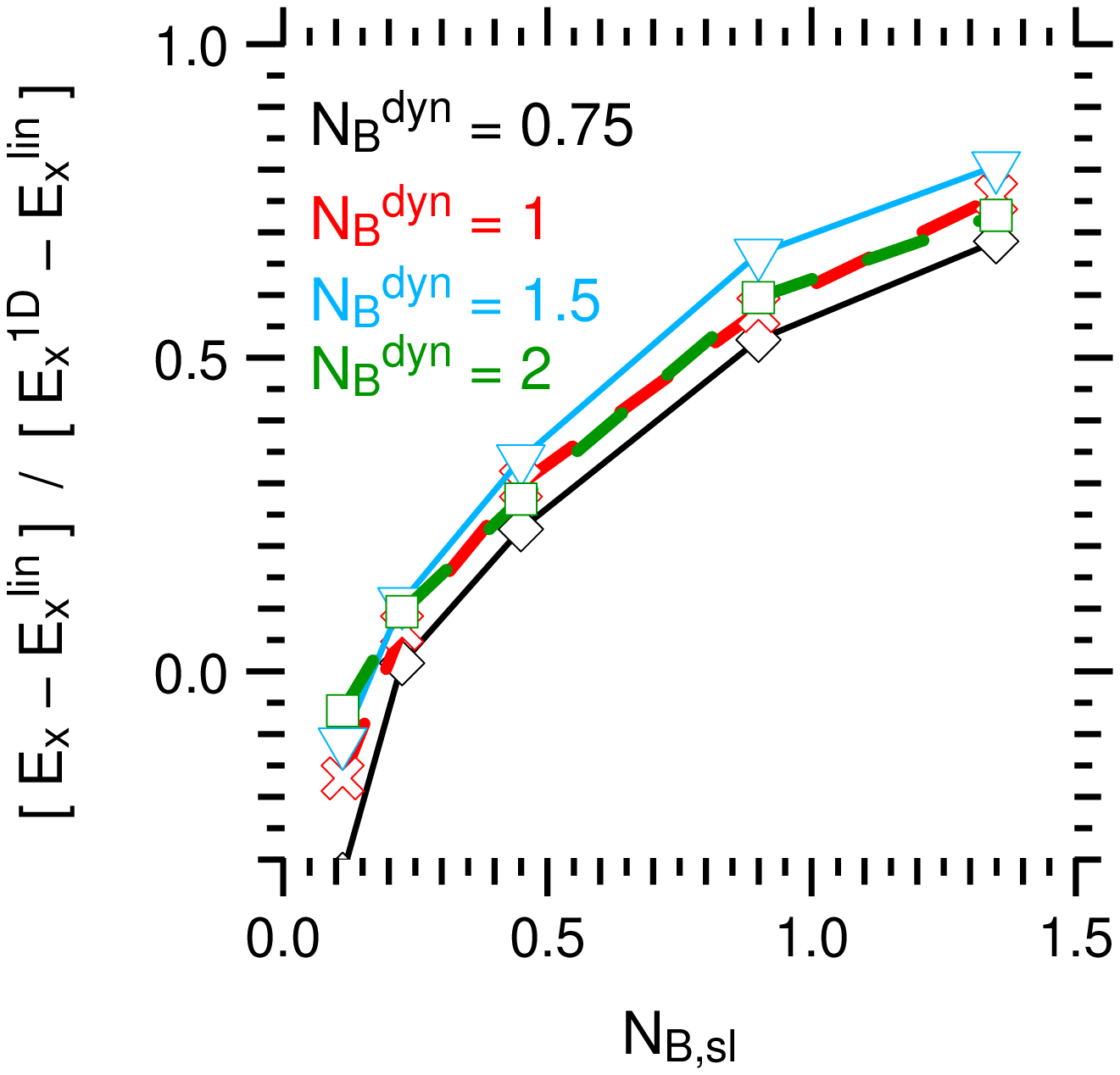}
  \caption{\colonl\ Departure from linear response for 2D \loki\
    runs, scaled to the same quantity from the 1D run \texttt{1D.35c}. The colored
    curves are taken at times when the dynamic bounce number $\NBx$ has reached
    the value indicated by the colored text. $\NBsl$ is found from linear response, using Eq.\ (\ref{eq:NBsllin}).}
  \label{fig:delexlin_Ly}
\end{figure}

The field magnitude $E_x(y=0)$ is plotted vs.\ the dynamic
bounce number $\NBx$ found using $E_x^0$ for the 2D
runs in Fig.\ \ref{fig:exlin_Ly}. The black curve is the analogous 1D run \texttt{1D.35c}. For
$\NBx \lesssim4$ there is a continuous increase in the response with profile
width $L_y$. This allows us to quantify trapping nonlinearity vs.\ $L_y$, which
we do in Fig.\ \ref{fig:delexlin_Ly}. The abscissa in that figure is the side
loss bounce number, $\NBsl$, computed with linear response as in Eq.\
(\ref{eq:NBsllin}). The ordinate is the field enhancement due to trapping,
scaled to the same quantity for the 1D run. This is shown at times corresponding
to several values of $\NBx$ ranging from 0.75 to 2. These times are early enough that the amplitudes have been
mostly increasing, with little oscillation due to the frequency shift.  The
curves agree well, and demonstrate the continuous development of trapping
effects with wide profiles.  Slightly more than half the 1D trapping effect
obtains for $\NBsl=1$, which vindicates our $N_B\sim1$ approximate threshold for
trapping.

The plasma response to a driver with transverse profile $h(y)$ differs from the
1D case. This can be seen in the ordinate of Fig.\ \ref{fig:delexlin_Ly} falling
below zero for the smallest $L_y=100$. There have been several linear
calculations of transit-time damping in LWs of finite extent, mostly by
integration along particle orbits \cite{short-transtime1-pop-1998,
  skjaeraasen-transtime-pop-1999}. Ref.\ \onlinecite{short-transtime1-pop-1998}
showed that, for a potential with a step-function profile in space, the
transit-time damping exceeds that for an infinite plane-wave for
$\om/k\vte\gg1$, while for $\om/k\vte \sim 1$ it can be less.  We adopt the
alternative approach of writing the response as a superposition of responses to
the Fourier modes comprising the drive. This is particularly convenient for our
$h(y)$, which (when periodically repeated) is composed of only two Fourier
modes. For simplicity we present the result for $h(y)$ periodically repeated,
instead of the actual \loki\ profile with compact support over $|y|<L_y/4$. The
compact case would lead to a continuous Fourier transform rather than discrete
series, and introduce a line width around the dominant modes. This does not
change the qualitative result. Unlike Ref.\
\onlinecite{short-transtime1-pop-1998}, our compact profile $h(y)$ is not a step
function but smooth, with $h$ and $h'$ continuous at all points (although $h''$
is not).

The drive $E_d$, made periodic in $y$, is
\begin{equation}
  \label{eq:17}
  E_d = {E_0\over4}e^{i(k_0x-\om_0t)}\lb 1 + {1\over2}e^{ik_1y} + {1\over2}e^{-ik_1y} \rb + c.c.
\end{equation}
A standard kinetic calculation, accounting for the fact that $\vec E_d$ has
no $y$ component and thus does not come from a potential, gives the field at $y=0$:
\begin{eqnarray}
  \label{eq:16}
  \Exlin(x,t,y=0) = E_0|R|\cos(k_0x-\om_0t+\al), \\
  2R = {1 \over 1+\chi_0} + {1+(1+(k_0/k_1)^2)^{-1} \chi_+ \over 1+\chi_+}.
\end{eqnarray}
Note that the linear $E_y(y=0)=0$ for our $\vec E_d$. $\al$ is a real
phase. $\chi$ is the collisionless susceptibility for $\nu_K=0$ from Eq.\ \ref{eq:chi}, which depends only on $\om$
and $k=|\vec k|$. $\chi_0=\chi(k_0,\om_0)$ and $\chi_+=\chi(k_+,\om_0)$ with
$k_+=(k_0^2+k_1^2)^{1/2}$. For $k_1=0$, we recover the 1D result Eq.\
(\ref{eq:ET}). Physically, the higher-$k$ modes induced by the transverse
profile are more Landau damped (as well as being slightly off resonance for the
fixed $\om_0$), which reduces the response. For the parameters of Table
\ref{tab:2D}, we find $|R|/|R|_{1D}=(0.801, 0.948)$ for $L_y=(100,200)$ where
$|R|_{1D}=11.9$ is the value for $L_y\rightarrow\infty$. We obtain a slight
decrease in the linear response for our sharpest profile ($L_y=100$), and an insignificant change
for wider ones. This is borne out by Fig.\ \ref{fig:exlin_Ly}. The red curve for
$L_y=100$ shows no signs of trapping, and reaches a steady level slightly more than
0.8 times the 1D linear value. The blue curve ($L_y=200$) shows a slight
trapping enhancement, and reaches a steady level slightly above 1.2x linear after
about 2 bounce periods.

\section{Coulomb Collisions} \label{s:coll} 

Collisions remove electrons from the trapping region via pitch-angle scattering
(from electron-ion and electron-electron collisions) as well as parallel drag
and diffusion (from only electron-electron collisions since $m_i/m_e\gg1$). We
adopt a Fokker-Planck collision operator, and discuss its validity in the Appendix:
\begin{eqnarray} \label{eq:fp}
  \p_tf &=& \nu_0(1+\zeff)u^{-3}\p_\mu\lb(1-\mu^2)\p_\mu f\rb \nonumber \\
   && + 2\nu_0u^{-2}\p_u(f+u^{-1}\p_uf) .
\end{eqnarray}
$\mu=\cos\ta$ where $\ta$ is the pitch angle between $\vec u$ and the $u_x$
direction, and $u=|\vec u|$. $\nu_0$ is a thermal electron-electron collision rate:
\begin{equation} \label{eq:nu0}
  \nu_0 \equiv {\omp \ln\La_{ee} \over 8\pi\NDe}.
\end{equation}
$\NDe=n_e\lde^3$ and $\ln\La_{ee}=24-\ln(n_e^{1/2}/T_e)$ ($n_e$ in cm$^{-3}$,
$T_e$ in eV) is the electron-electron Coulomb logarithm appropriate for $T_e>10$
eV (Ref.\ \onlinecite{nrl-formulary}, p.\ 34). The effective charge state is
\begin{equation}
\zeff \equiv \sum_i {f_iZ_i^2\over \bar Z}{\ln\La_{ei} \over \ln\La_{ee}},
\end{equation}
where $n_I=\sum_in_i$ is the total ion density, $\bar Z = \sum_iZ_if_i$ with
$f_i=n_i/n_I$; $\sum_if_i=1$, and $\ln\La_{ei}$ is the electron-ion Coulomb
logarithm \cite{nrl-formulary}.

In section \ref{s:icf} we apply our results to Langmuir waves generated by Raman
scattering in underdense ICF plasmas, which are typically low-Z. For instance,
NIF ignition hohlraum designs currently use an He gas fill (with H/He mixtures
contemplated), and plastic ablators (57\% H, 42\% C atomic fractions). This gives $\zeff=5.08$ when
fully-ionized and $\ln\La_{ei}=\ln\La_{ee}$. Be and
diamond ablators are also being considered. For illustration, we take $\zeff=1$ as the lowest
reasonable value (fully-ionized H), and use $\zeff=4$ (fully-ionized Be) to
represent an ablator plasma.

It is useful to define a unitless time $\hat t$ (different from the side loss $\hat t$ used above), which demonstrates some of the basic collisional scaling:
\begin{eqnarray} \label{eq:thatcoll}
  \hat t &\equiv& {\nu_ct \over \de N},  \\ 
  \nu_c &\equiv& {\pi^2 \over 16}{\nu_{0} \over u_p^3}(k\la_{De})^2 = 
  {\pi\over128}{(k\lde)^5\over(\om/\omp)^3} {\ln\La_{ee}\over N_{De}} \omp. \label{eq:nuccoll}
\end{eqnarray}
Our collisional calculation of the trapped fraction is detailed in the Appendix. The key
observation is that the distribution in the trapping region can be decomposed
into Fourier modes $\sin[n\pi((v_x-v_p)/v_{tr}+1/2)]$ for $n=1,3,...$, and the
diffusion rate of mode $n$ is proportional to $n^2$. After a short time, only
electrons in the $n=1$ mode remain trapped, so it suffices to consider just the
number in the $n=1$ mode. At $t=0$, this is 81\% of the total (the other 19\%
rapidly diffuses out). The upshot is that
$\ntrc$, the fraction of initially trapped particles remaining in the
fundamental mode after time $t$, is
\begin{equation} \label{eq:Ntrco}
  \ntrc(\hat t,\zeff,u_p) = 0.81\int_0^\infty d\upe  \upe  \exp\lb -\upe ^2/2 -D\hat t \rb.
\end{equation}
$D(\upe ,u_p,\zeff)$ is given in Eq.\ (\ref{eq:D}). 

\begin{figure}
  \centering
  \includegraphics[width=3.3in]{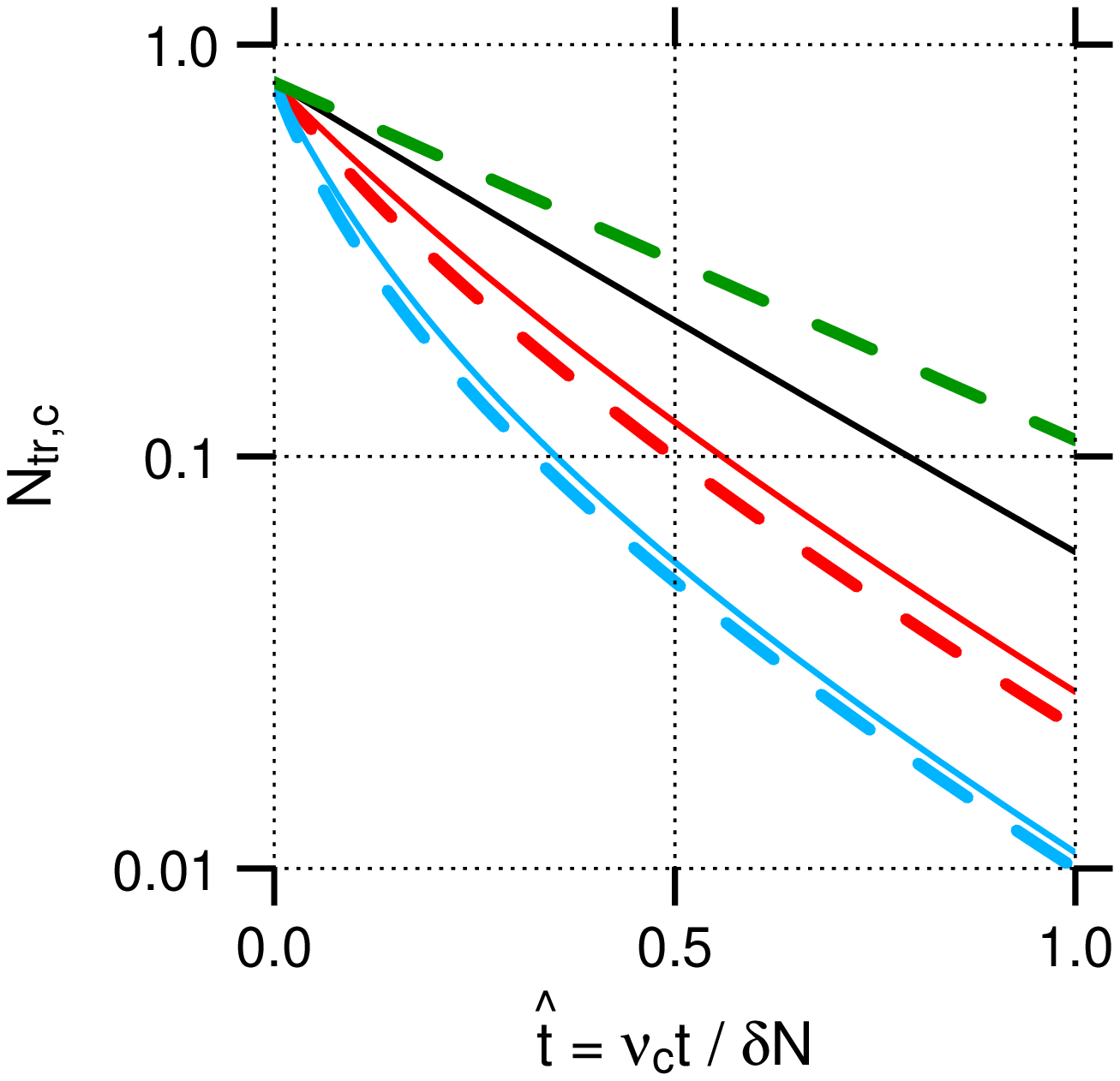}
  \caption{\colonl\ Trapped fraction due to collisions $\ntrc$ vs.\ unitless
    time $\hat t$ defined in Eq.\ (\ref{eq:thatcoll}). Solid curves are exact results
    from Eq.\ (\ref{eq:Ntrco}). $(u_p,\zeff)$ = (2,1), (4,1), and (4,4) for black,
    red, and blue, respectively. Dashed red and blue curves are approximate results for
    $u_p\rightarrow\infty$ from Eq.\
    (\ref{eq:Ntrc_upinf}), which depend only on $\zeff$ and not $u_p$. Green
    dashed curve is $0.81\exp[-2\hat t]$, the approximate form neglecting the term
    proportional to $\hat t$ in the denominator. $\ntrc(t=0)=0.81$ and not unity due to
    electrons not initially in the fundamental $u_x$ mode.}
  \label{fig:ntrc}
\end{figure}

Eq.\ (\ref{eq:Ntrco}) is an implicit, integral equation for $\hat t$ as a
function of $\zeff$, $u_p$, and $\ntrc$. We find the ``exact'' solution by
performing the integral numerically, and interpolating $\hat t$ for a desired
$\ntrc$. We derive an approximate solution, valid for $u_p\gg1$, for $\hat t$
in the Appendix. The result is
\begin{equation}
  \label{eq:thatappr}
  \hat t \approx \hat t_0 + \hat t_1u_p^{-2}.
\end{equation}
$\hat t_0$ and $\hat t_1$ are both positive and depend only on $\zeff$, so $\hat
t$ decreases with
increasing $u_p$. Figure \ref{fig:ntrc} plots $\ntrc(\hat t)$ for several $u_p$ and $\zeff$, using the exact results (solid curves) and the approximate form for
$u_p\rightarrow\infty$ of Eq.\ (\ref{eq:Ntrc_upinf}) (dashed curves).  Few
electrons remain trapped at $\hat t=1$. The approximate forms are quite good,
even though $u_p$ is not that large.

\begin{figure}
  \centering
  \includegraphics[width=3.3in]{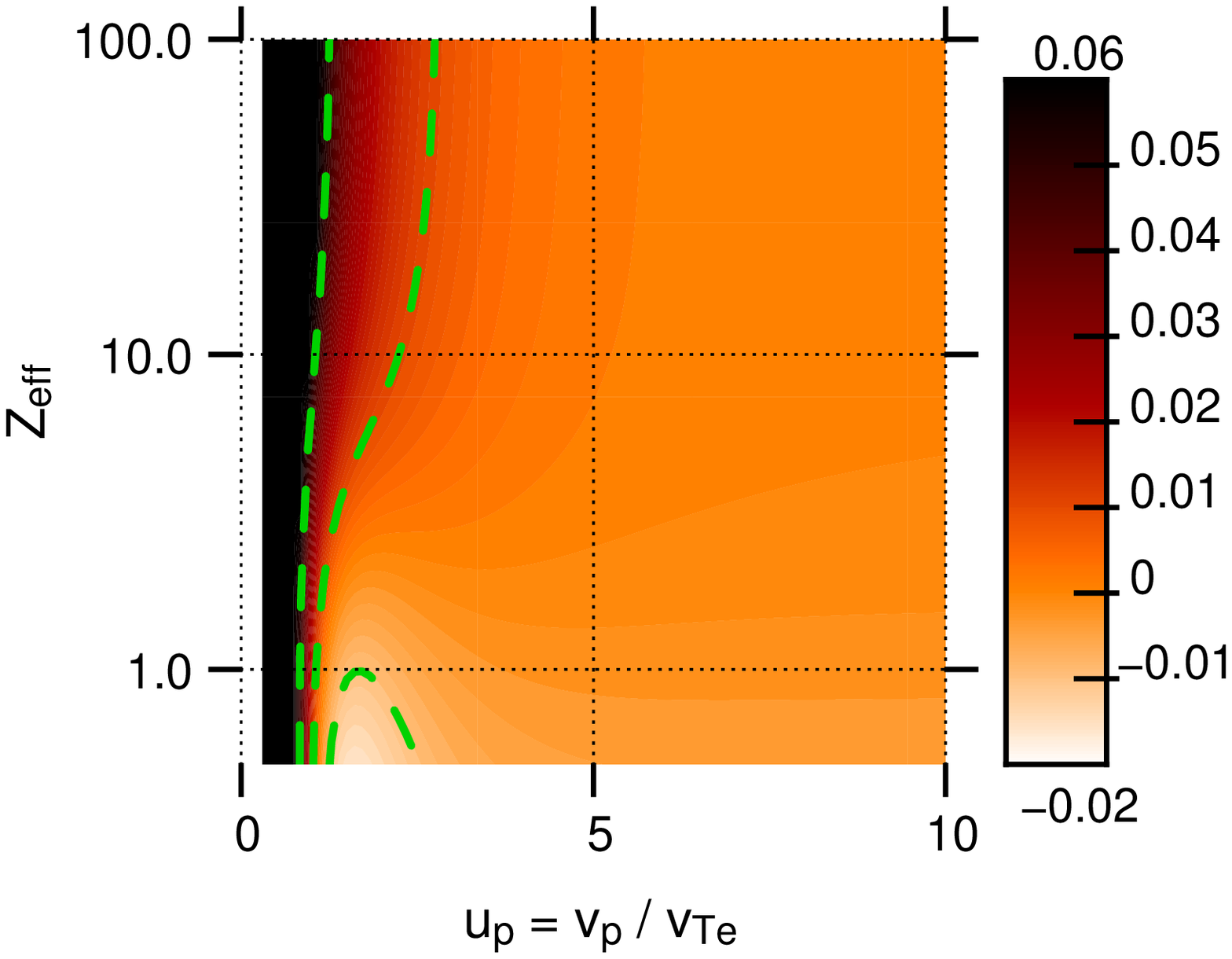}
  \caption{\colonl\ Relative error $\ep=1-\thatappr/\thatex$ between the exact
    and approximate results from Eqs.\ (\ref{eq:Ntrco},\ref{eq:thatappr}), respectively, for collisional $\hat t$; see text for details. The green curves are $\ep=[-0.01,0.01,0.05]$.}
  \label{fig:thatcollerr}
\end{figure}

Figure \ref{fig:thatcollerr} displays the relative error
$\ep\equiv1-\thatappr/\thatex$ between $\hat t$ for $\ntrc=1/2$ computed two
ways. The exact $\thatex$ is found numerically, and $\thatappr$ is from
Eq. (\ref{eq:thatappr}), with Eq.\ (\ref{eq:t0quad}) for $\hat t_0$ and  Eq.\ (\ref{eq:t1}) for $\hat t_1$. The
agreement is excellent, within 1\% for most of parameter space.

The collisional detrapping rate $\nudc$ is
\begin{equation}
  \nudc = {\nu_c \over \de N}{\ln(1/\ntr) \over \hat t}.
\end{equation}
Note that $\nudc\sim \de N^{-1}$
since $u_{tr} \sim \de N^{1/2}$: the larger the wave
amplitude, the wider the trapping region extends in velocity, and collisions take
longer to remove the electron velocity from this region. Recall that $\nudc$
depends slightly on the choice of $\ntr$ due to the non-exponential decay of
$\ntr$ with $\hat t$; as with convective loss we choose $\ntr=1/2$.

The collisional bounce number is
\begin{equation}
  N_{B,c} = \lb{\de N\over\de N_c}\rb^{3/2}, \qquad 
  \de N_c = \lb 2\pi {\nu_c \over \omega_{pe}} {\ln(1/\ntr) \over \hat t} \rb^{2/3}. 
\end{equation}
The amplitude exponent for collisions is $p_c=3/2$, unlike the convective loss
value of 1/2.  This stems from the fact that $\nu_d$ for collisions is
amplitude-dependent while for convective loss it is not. We now construct the
overall bounce number $N_{B,O}$ for convective side loss and collisions, as
outlined above.  Assuming that separate detrapping processes are independent,
and their detrapping rates add, yields
\begin{equation} \label{e:NBO}
  N_{B,O}^{-1} = \NBsl^{-1}+N_{B,c}^{-1} = \lb{\dNsl\over \de N}\rb^{1/2} + \lb{\de N_c\over \de N}\rb^{3/2}.
\end{equation}
We define an overall threshold amplitude $\de N_O$ such that $N_{B,O}[\de N=\de N_O]=1$. Eq.\ (\ref{e:NBO}) gives a cubic equation for $a\equiv \de N_O^{1/2}$:
\begin{equation}
  a^3 - \dNsl^{1/2}a^2 - \de N_c^{3/2} = 0.
\end{equation}

There are two ways to compare the relative importance of side loss and
collisions. One is: for which process must the wave amplitude $\de N$ be larger
for trapping to be significant ($N_B=1$)? The other is: for a given $\de N$,
which process will detrap more effectively? The two views are not equivalent,
due to the different dependence of the side loss and collisional detrapping rate
on $\de N$. The first amounts to comparing the thresholds $\dNsl$ and $\dNc$,
which can be computed just from plasma and wave properties without knowing
$\de N$. The ratio of detrapping rates can be written in terms of a
critical amplitude $\dNcr$:
\begin{equation}
  \label{eq:dNcr}
  {\nudc \over \nudsl} = {\dNcr \over \de N}, \qquad 
  \dNcr \equiv {\ln2 \over \hat t \Ksl }{\nu_c \over \omp}{L_\perp \over \lde}.
\end{equation}

\section{Parameter study for ICF underdense plasmas} \label{s:icf} 

We now apply our analysis to ICF conditions where stimulated Raman scattering
(SRS) can occur, namely the underdense coronal plasma. SRS is a parametric
three-wave process where a pump light wave such as a laser (we which label mode
0) decays to a scattered light wave (mode 1) and a Langmuir wave (mode 2). We
restrict ourselves to exact backscatter (SRBS; $\vec k_1$ anti-parallel to $\vec
k_0$), as this generates the largest $k_2$ (smallest $v_{p2}/\vte$) and thus
makes trapping effects more important (small transverse components to $\vec k_2$
have little effect on the phase velocity). Both measurements and
simulations with the paraxial-envelope propagation code \ftd\ \cite{berger-f3d-pop-1998} have shown backscatter to be the dominant
direction for SRS. With $\vec k_i=k_i\hat z$, the phase-matching conditions are
$\om_0=\om_1+\om_2$ and $k_0=k_1+k_2$ with $k_1<0$. We employ the (cold)
light-wave dispersion relation $\om_i^2=(ck_i)^2+\omp^2$ for modes $i=$0 and 1,
and use the vacuum wavelength $\la_i=2\pi c/\om_i$. Frequency matching thus
requires $n_e<n_{cr}/4$, with $n_{cr,i} \equiv (\ep_0m_e/e^2)\om_i^2$ the
critical density for mode $i$, and $n_{cr}=n_{cr,0}$. For specific examples we
choose $\la_0$= 351 nm, appropriate for frequency-tripled UV light currently in
use on NIF. Specific plasma conditions thought to be typical for
SRBS on NIF ignition targets, during early to mid peak laser power, are
$n_e/n_{cr}=0.1$ and $T_e=2$ keV ($\la_1\approx 550$ nm)
\cite{strozzi-srs-dpp-2011}. The scattered wavelength continuously increases
during a NIF experiment, consistent with the hohlraum filling to higher density.

An important case for this paper is LW's driven by SRBS in the speckles of a
phase-plate-smoothed laser beam \cite{kato-rpp-prl-1984}. For a laser wavelength
$\la_0$ and square RPP with optics F-number $F$, the intense speckles have
$L_\perp \approx F\la_0$ and $L_{||} \approx 5F^2\la_0$ (see Ref.\
\onlinecite{garnier-rpp-pop-2001}).  A speckled beam is not the only situation
where SRS can occur; for instance, there has been recent interest in
re-amplification of backscatter by crossing laser beams
\cite{kirkwood-multibeam-pop-2011} and backward Raman amplifiers
\cite{yampolsky-ramanamp-pop-2011}. However, for a single laser beam, experiments at Omega and \ftd\
simulations show speckle physics, and its modification
by beam smoothing, must be accounted for to accurately model SRS
\cite{froula-srs-prl-2009, froula-lpi-pop-2010}.  Experiments have also verified
the increase in backscatter with increased gain per speckle length, by changing
the laser aperture and thus the effective $F$ \cite{froula-privcom-Fnum-2009}. We therefore focus on speckles. On NIF, four laser beams, each
smoothed by a phase plate and with an overall $F=22$ square aperture, are
grouped into a ``quad'' which yields an effective square aperture of
$F\approx8$. We thus use $F=8$ for illustration. As the beams of a quad
propagate through a target, they can separate from one another, refract, and
undergo other effects that change the shape of their effective aperture and
speckle pattern. We do not pursue this further here, but it should be born in
mind when applying our analysis. Also the ratio $L_\perp/L_{||}=1/5F$ is so
small that $\nudel/\nudsl \approx (\Kel/\Ksl)u_p/5F=0.0083u_p$ (3D) is small for
essentially all speckles of interest. Thus side loss is a more potent detrapping
mechanism than end loss, in speckles.

\begin{figure}
  \centering
  \includegraphics[width=3.3in]{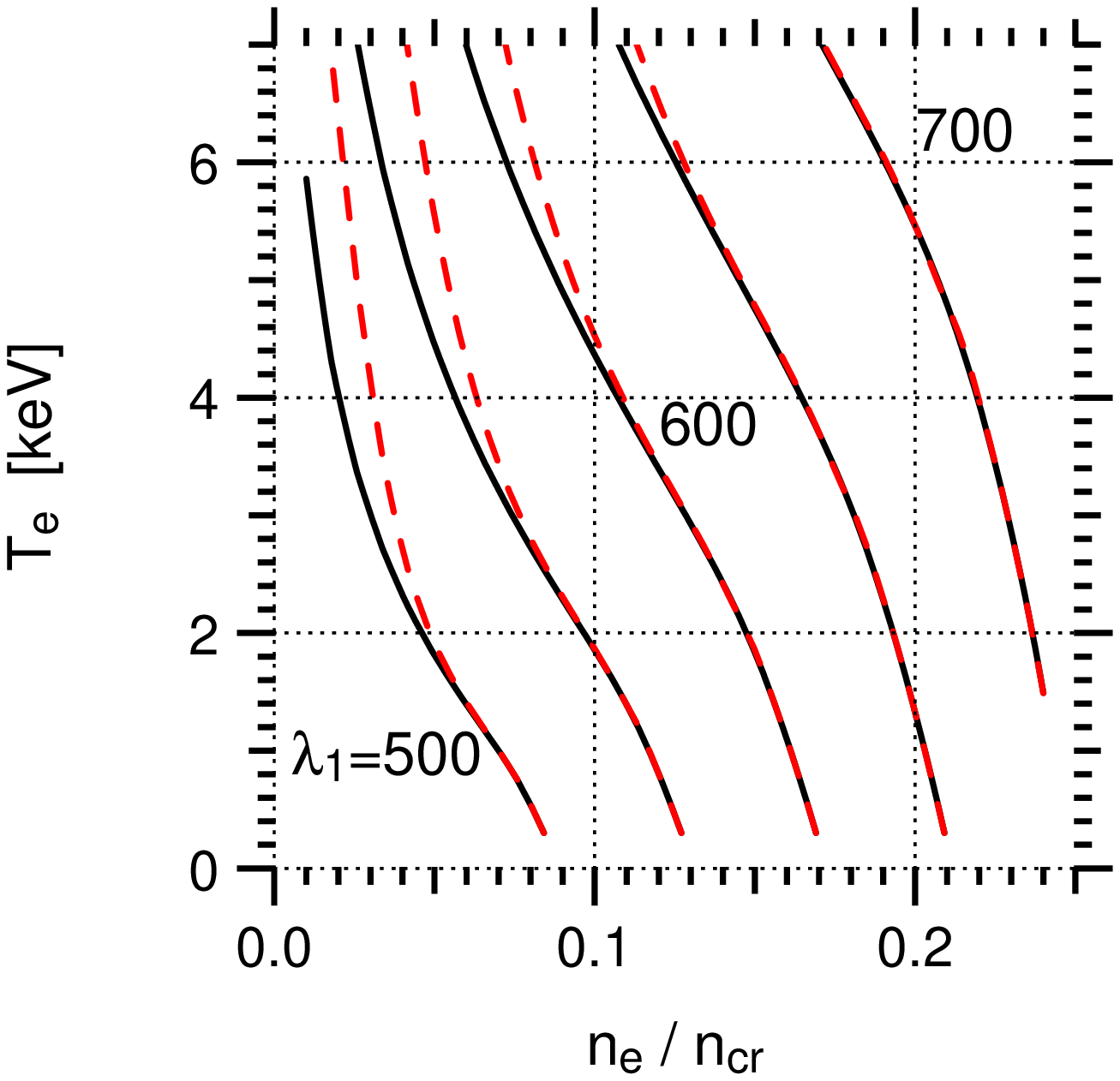}
  \caption{\colonl\ Wavelength $\la_1$ in nm for SRBS light (increments of 50
    nm), for a pump wavelength $\la_0$= 351 nm. Black solid: $\la_1$
    phase-matched with a natural Langmuir wave, satisfying the dispersion
    relation Eq.\ (\ref{eq:chip1}) with $\nu_K=0$. Red dash: $\la_1$ for the maximum local SRBS spatial gain rate.}
  \label{fig:lam1}
\end{figure}

\begin{figure}
  \centering
  \includegraphics[width=3.3in]{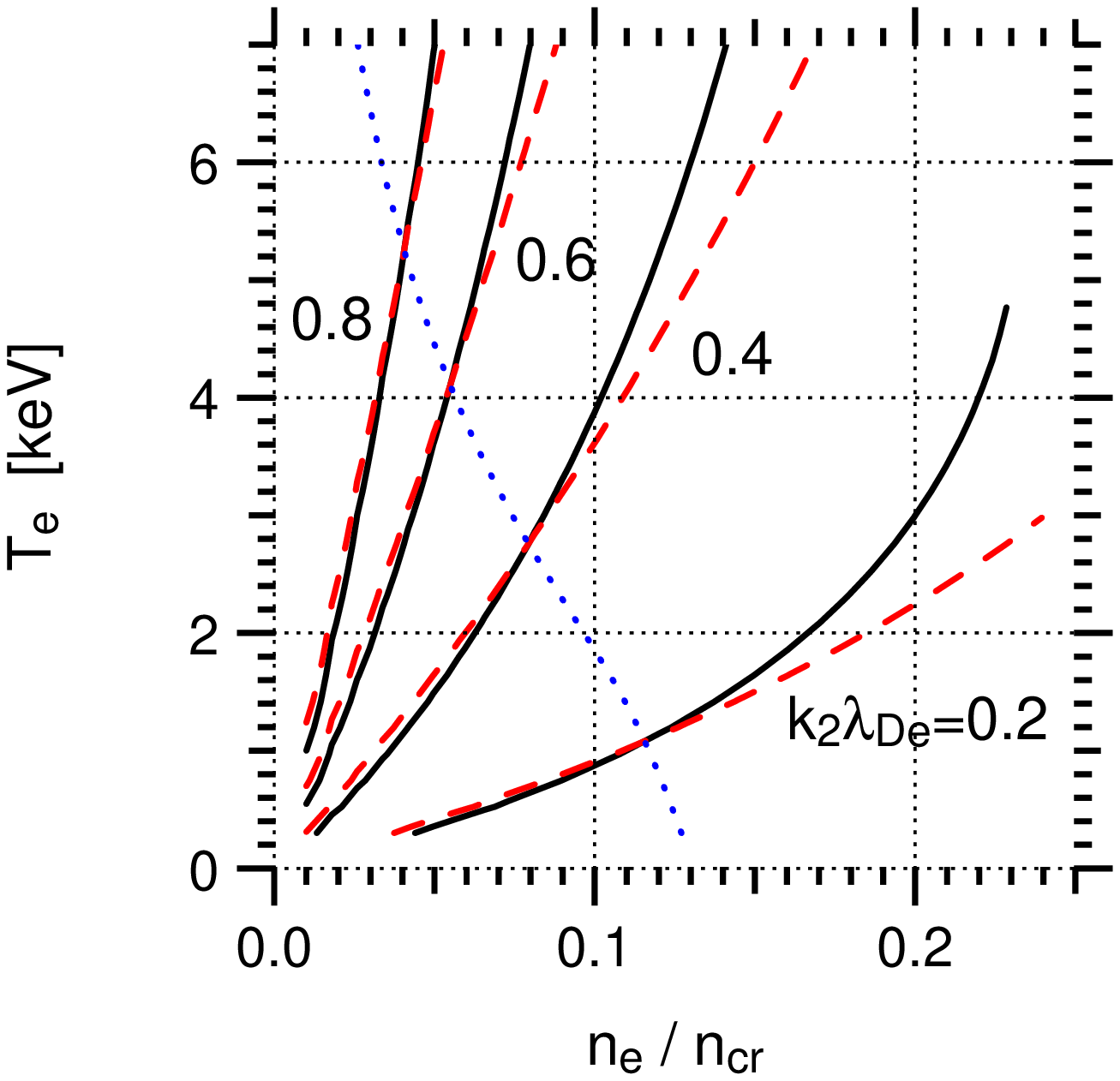}
  \caption{\colonl\ Langmuir wave $k_2\lde$ phase-matched for SRBS with
    different $\la_1$ choices, for a pump wavelength $\la_0$= 351 nm. Black
    solid: $\la_1$ phase-matched with a natural Langmuir wave, as in black solid
    curves of Fig.\ \ref{fig:lam1}. Red dash: $\la_1$=550 nm ($\omega$'s fixed,
    $k$'s vary). Blue dot: $\la_1$=550 nm black-solid contour from Fig.\ \ref{fig:lam1}.}
  \label{fig:k2lde}
\end{figure}

To quantify detrapping rates, we consider the threshold amplitudes $\dNsl$ and
$\de N_c$. Unlike $\dNsl$, $\de N_c$ depends on $\om_2$ and $k_2$ of the
Langmuir wave. For a given set of plasma conditions, the choice of $(\om_2,k_2)$
is not unique but depends on the application. For SRS developing locally, one
can choose the LW corresponding to the largest growth rate for those
conditions. Another approach is to consider a single scattered-light frequency
as it propagates through a target. We consider only $k$ variations
induced by spatial profiles and not $\om$ variations due to temporal plasma
evolution \cite{dewandre-wshift-pof-1981} (which is mostly relevant to stimulated Brillouin scattering). In this case, the matching conditions given the local plasma properties dictate how $k_2$ varies.

Figure \ref{fig:lam1} presents the local $\la_1$ for SRBS computed in two
ways. The black curves are found by phase-matching with a ``natural'' LW, by
which we mean $\om_2=\mathrm{Re}[\om_{2c}]$ where complex $\om_{2c}$ satisfies
\begin{equation} \label{eq:chip1}
1+\chi[k_{2r},\omega_{2c}]=0
\end{equation}
with real $k_{2r}=k_0-k_1$. To find
$\om_{2c}$, we set $\nu_K=0$ and recover the usual collisionless $\chi$. We use
$\nu_K\neq0$ below as a simple way to include collisional LW damping when Landau
damping is negligible. The red curves in Fig.\ \ref{fig:lam1} are the $\la_1$
which maximizes the local spatial SRBS gain rate in the strong damping limit \cite{strozzi-dep-pop-2008}:
\begin{equation}
  \label{eq:gainrate}
  \p_z\ln i_1(\lambda_1,z) = \lb -{2\pi r_e \over m_ec^2}{I_0\over \omega_0k_0} \rb \lb
  {k_2^2 \over |k_1|} \mathrm{Im}{\chi \over 1+\chi} \rb.
\end{equation}
We use the collisionless $\chi$ with $\nu_K=0$. The first bracket is independent
of $\lambda_1$, while the second bracket is not. The two results for $\lambda_1$
in Fig.\ \ref{fig:lam1} are very close except for high-$k_2\lde$ LW's (low
$n_e$, high $T_e$), where Landau damping and its variation with $\lambda_1$ is
significant. We choose for convenience to use $\la_1$ matched to a natural LW
below. We display in Fig.\ \ref{fig:k2lde} the $k_2\lde$ corresponding to two
choices of $\la_1$. The black curves use the $\la_1$ phase-matched to a natural
LW (the black curves in Fig.\ \ref{fig:lam1}), while the red curves are for a
constant $\la_1$= 550 nm.

\begin{figure}
  \centering
  \includegraphics[width=3.3in]{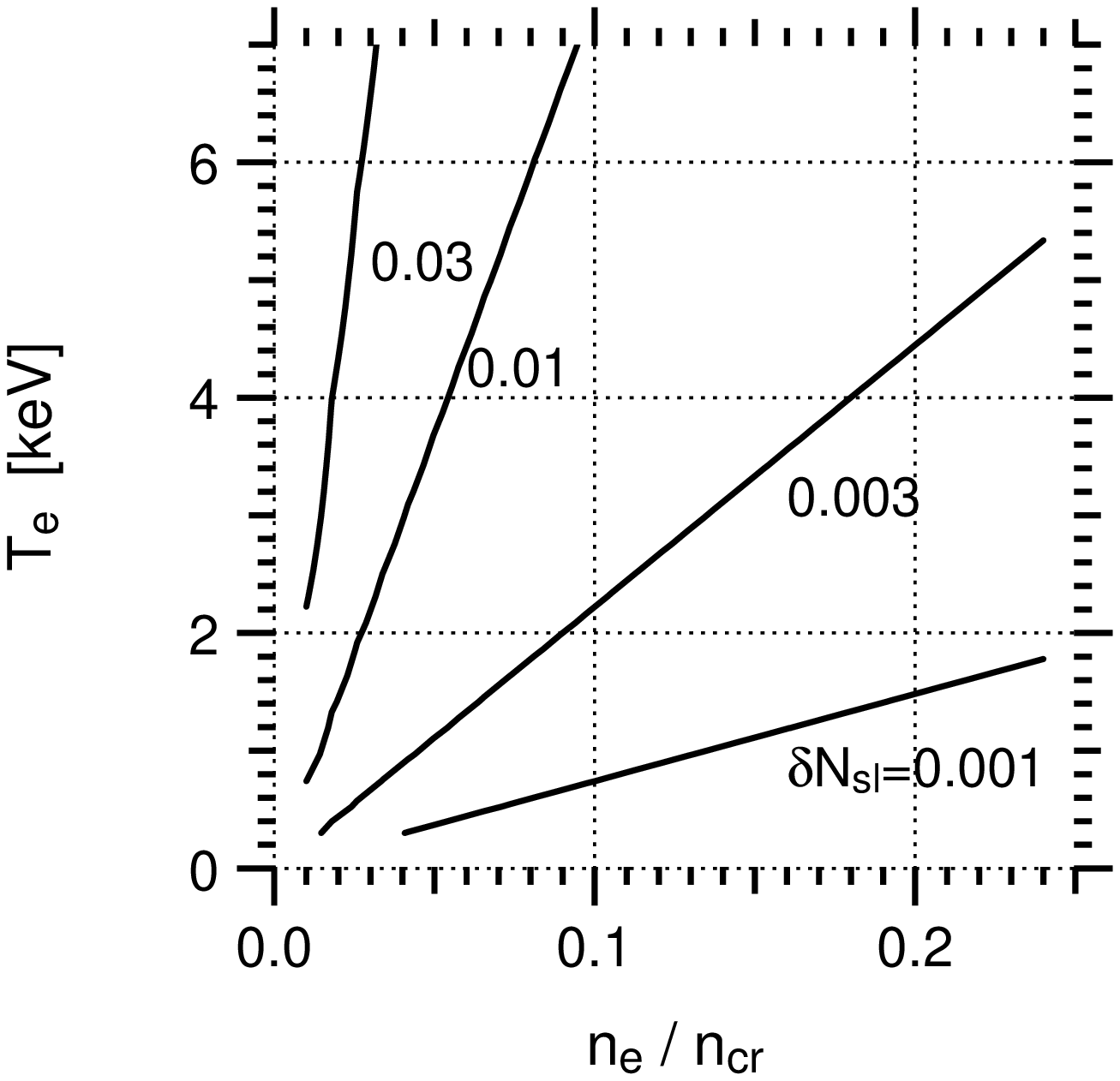}
  \caption{Threshold Langmuir wave amplitude for side loss, $\dNsl$, for $L_\perp$=8$\times$351 nm.}
  \label{fig:dNsl}
\end{figure}

\begin{figure}
  \centering
  \includegraphics[width=3.3in]{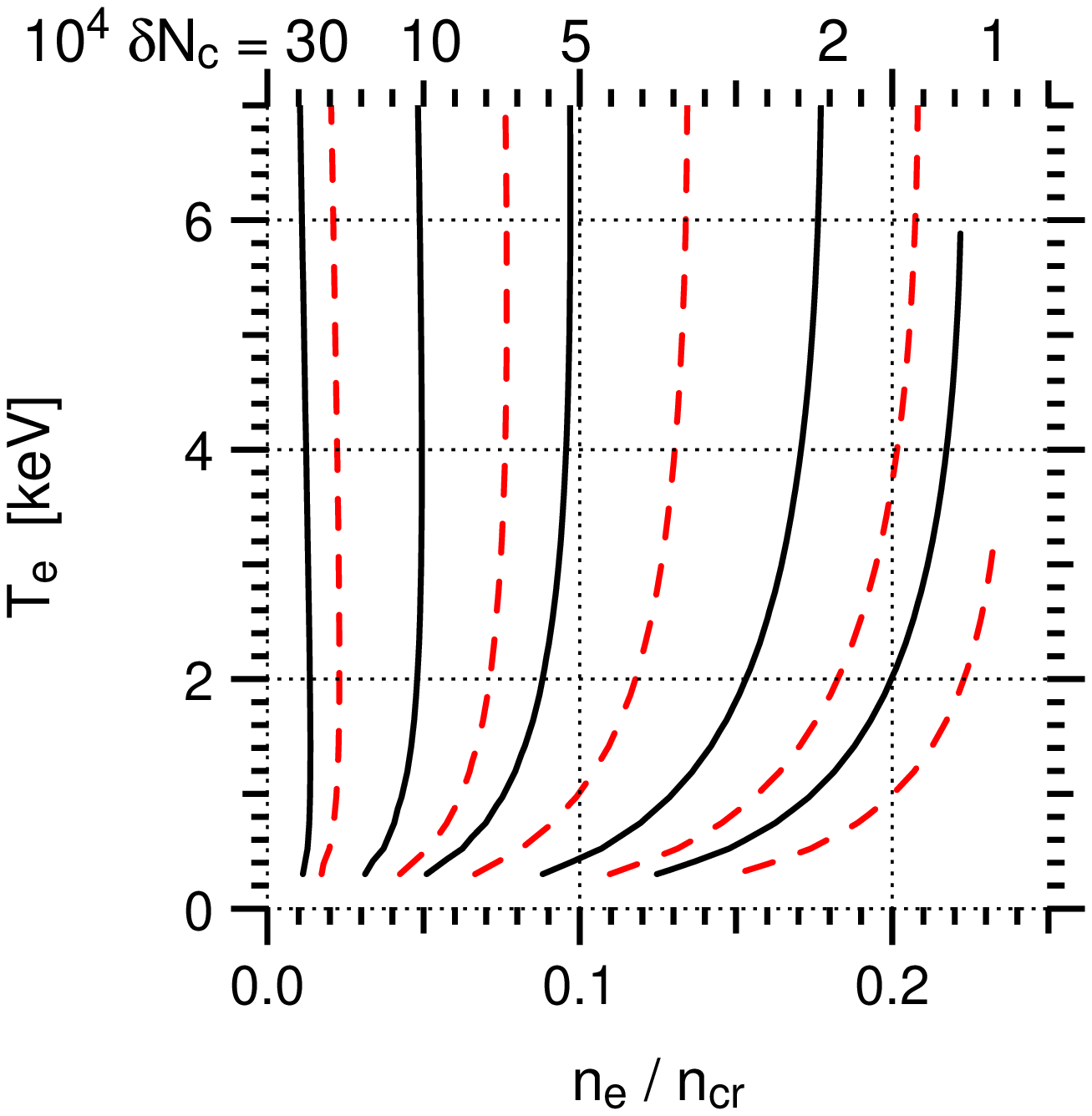}
  \caption{\colonl\ Threshold Langmuir wave amplitude for collisions, $\de
    N_c$. LW $\om_2$ and $k_2$ are for phase-matched SRBS with a natural LW
    (black solid curves of Fig.\ \ref{fig:lam1}), and pump wavelength $\la_0$= 351 nm. Black solid: $\zeff$=1. Red dash: $\zeff$=4. From right to left, curves are for $\de N_c=[1,2,5,10,30]\times10^{-4}$.}
  \label{fig:dNcoll}
\end{figure}

\begin{figure}
  \centering
  \includegraphics[width=3.3in]{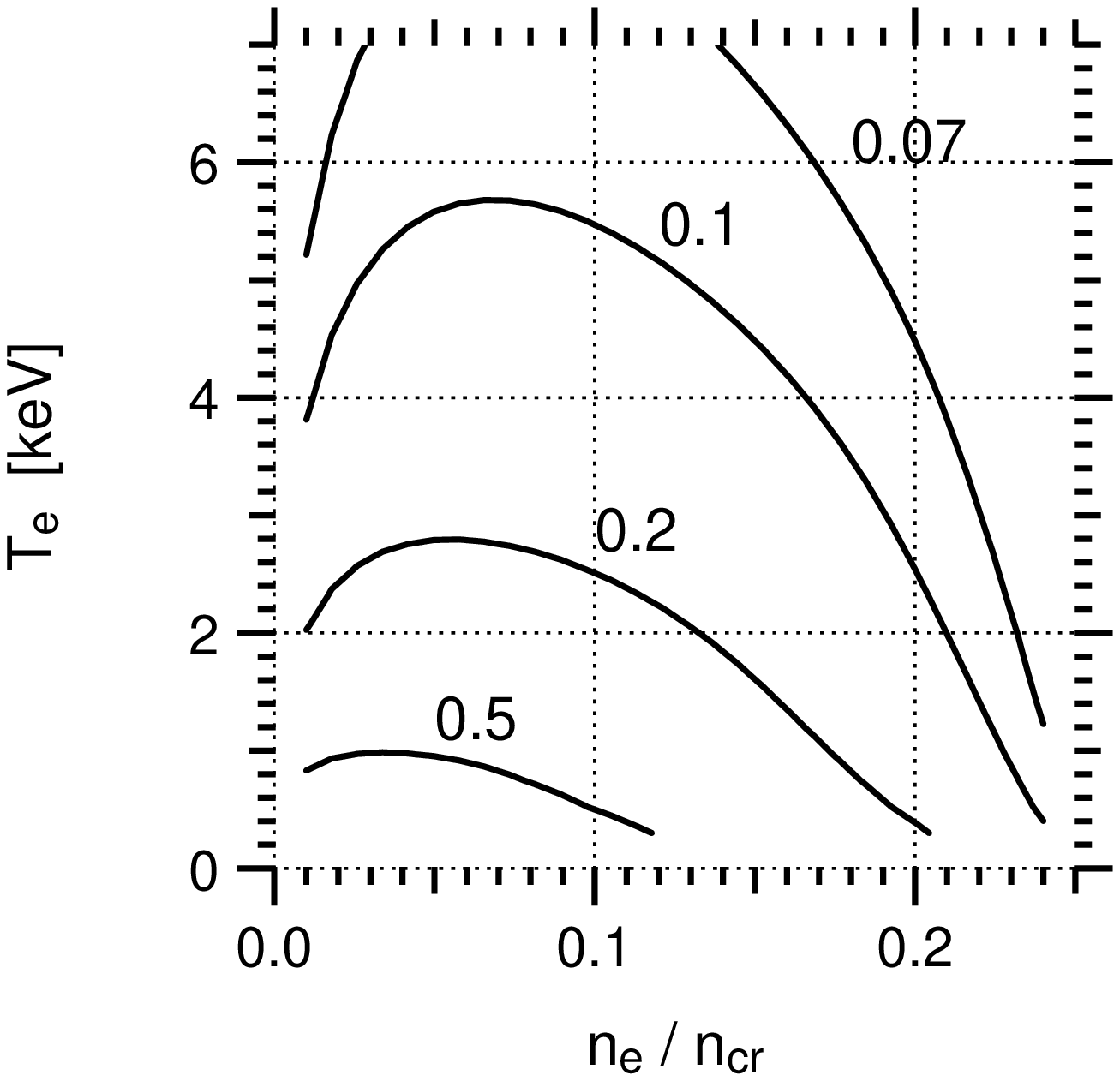}
  \caption{Ratio $\de N_c/\dNsl$ from Figs.\ \ref{fig:dNsl} and \ref{fig:dNcoll}
    for $\zeff$=4.}
  \label{fig:dNrat}
\end{figure}

\begin{figure}
  \centering
  \includegraphics[width=3.3in]{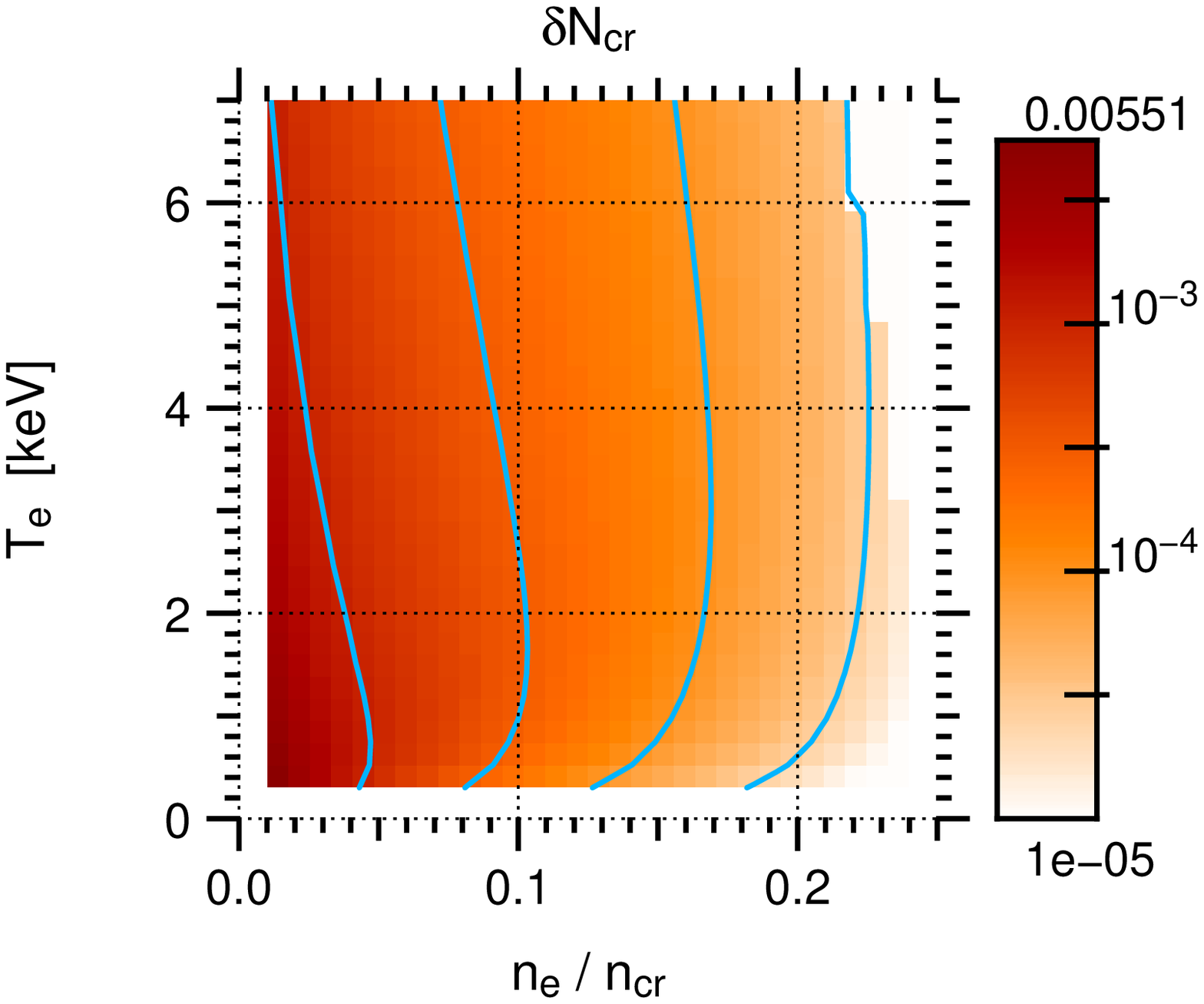}
  \caption{\colonl\ Critical amplitude $\dNcr$ from Eq.\ (\ref{eq:dNcr}) for
    $\zeff$=4. The side loss detrapping rate is faster than the collisional one
    for $\de N>\dNcr$. Blue curves are for $\dNcr = 3\times 10^{-5}, 1\times10^{-4}, 3\times10^{-4}, 1\times10^{-3}$.}
  \label{fig:dNcr}
\end{figure}

The side loss threshold $\dNsl$ is shown in Fig.\ \ref{fig:dNsl}, for
$L_\perp=F\la_0$ and $F=8$. It simply represents the variation in $\lde$, and is
independent of $(\om_2,k_2)$. Figure \ref{fig:dNcoll} depicts the
collisional threshold $\de N_c$ for $\zeff$= 1 and 4. The decrease of $\de N_c$
with electron density is mainly due to the decrease of the $(k\lde)^5$ factor in
$\nu_c$ (see Eq.\ (\ref{eq:nuccoll})), which in turn is due to the $1/v_p^3$
fall in the Coulomb cross-section (see Eq.\ (\ref{eq:nu0})). The ratio $\de N_c/\de
N_{sl}$ is displayed in Fig.\ \ref{fig:dNrat}, which indicates collisions have a
minor effect except for low $n_e$ and low $T_e$; this relative importance
depends strongly on the transverse length $L_\perp$ chosen for side loss. Figure
\ref{fig:dNcr} plots the critical amplitude $\dNcr$ from Eq.\
(\ref{eq:dNcr}). For $\de N>\dNcr$, the side loss detrapping rate exceeds the
collisional rate. $\dNcr$ is larger at smaller $n_e$, indicating collisional
detrapping is more relevant. 2D particle-in-cell simulations with the VPIC code
of Raman amplifier experiments \cite{kirkwood-ramanamp-jpp-2011, kirkwood-multibeam-pop-2011} found that collisions mattered for low-intensity seed light waves in a low-density plasma ($n_e/n_{cr}\sim0.01$).

The reflectivities which correspond to trapping nonlinearity can also be
estimated. We assume the LW's are in the strong damping limit \cite{strozzi-dep-pop-2008}, and write
\begin{equation}
  \label{eq:3}
  \de N = {1\over2} {(k_2\lde)^2 \over |1+\chi|} {V_0V_1 \over \vte^2}.
\end{equation}
$V_i=eE_i/m_e\om_i$ is the oscillation velocity for light wave $i$; in practical
quantities we have $(V_i/c)^2 = I_i\la_i^2/\eta_iP_0$, with $P_0 \equiv
2\pi^2(\ep_0/e^2)m_e^2c^5=1.37\times 10^{18}$ W cm$^{-2}\cdot\mu$m$^2$. $\eta_i=[1-n_e/n_{cr,i}]^{1/2}$ reflects the decrease in group velocity. With reflectivity $R=I_1/I_0$, we find
\begin{eqnarray}
  \label{eq:4}
  \de N &=& {I_0 \over I_{cr}}R^{1/2},  \\
  I_{cr} &\equiv& 2{|1+\chi| \over (k_2\lde)^2 }{T_e \over m_ec^2} { P_0 \over \la_0\la_1}(\eta_0\eta_1)^{-1/2}. \label{eq:Icr}
\end{eqnarray}
The ``critical intensity'' $I_{cr}$ is introduced for convenience. Since
trapping effects become significant for $\de N\approx\de N_i$ where $\de N_i$ is
the detrapping threshold for process $i$, we define the threshold reflectivity
$\Rthri$ for which $\de N=\de N_i$:
\begin{equation}
  \label{eq:Rthr}
  \Rthri = \lb {I_{cr} \over I_0} \de N_i \rb^2.
\end{equation}

\begin{figure}
  \centering
  \includegraphics[width=3.3in]{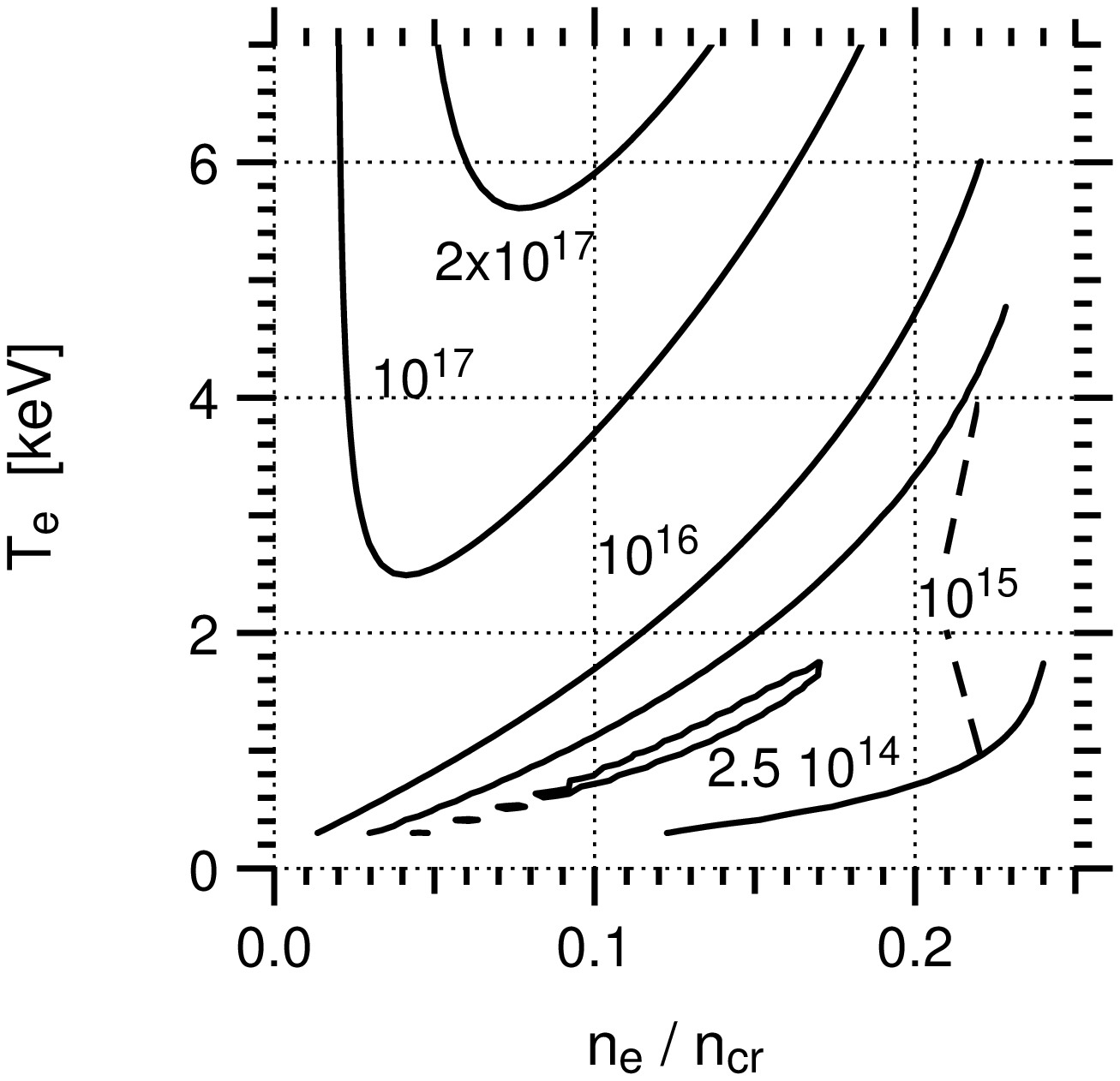}
  \caption{Critical intensity $I_{cr}$ from Eq.\ (\ref{eq:Icr}) in W/cm$^2$ relating laser intensity $I_0$ and reflectivity $R$ to LW amplitude: $\de N=(I_0/I_{cr})R^{1/2}$.}
  \label{fig:Icr}
\end{figure}

\begin{figure}
  \centering
  \includegraphics[width=3in]{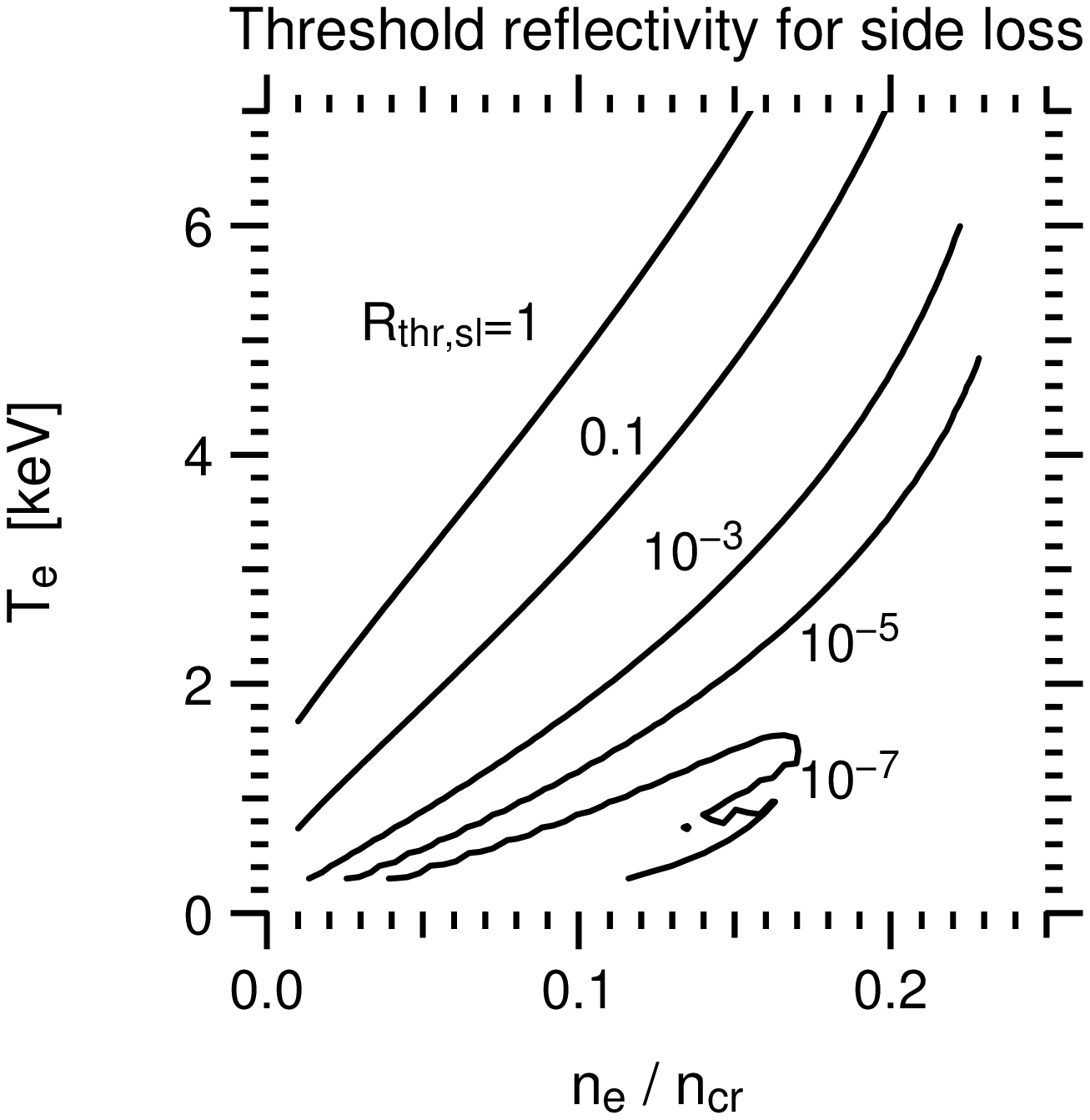}
  \caption{Threshold reflectivity $\Rthrsl \propto 1/I_0^2$ from Eq.\ (\ref{eq:Rthr}) to
    overcome side loss for $I_0=10^{15}$ W/cm$^2$.}
  \label{fig:Rthr}
\end{figure}

To illustrate the threshold reflectivity, we consider SRBS of the phase-matched natural LW. The critical intensity $I_{cr}$ is plotted in Fig.\ \ref{fig:Icr}. We use $\chi$ from Eq.\ (\ref{eq:chi}) including the Krook operator $\nu_K\neq0$ to damp the LW when $k_2\lde$ is small and Landau damping is ineffective. For this purpose we choose $\nu_K$ to be the collisional, unmagnetized frictional drag rate in the electron momentum equation \cite{braginskii-xport-1965, epperlein-xport-pof-1986}. This rate is appropriate for the drag on the bulk sloshing motion of the electrons in the LW electric field, and not collisions of resonant electrons with $v\approx v_p$. $I_{cr}$ minimizes near the lower-right corner near the $2.5\times10^{14}$ curve, where Landau and collisional damping are both weak. The threshold reflectivity $\Rthrsl$ to overcome side loss is plotted in Fig.\ \ref{fig:Rthr}, for a pump with $I_0=10^{15}$ W/cm$^2$. A small reflectivity produces a large LW in the lower-right corner where damping is weak, thus allowing trapping to more easily occur.

Our analysis assumes a Maxwellian electron distribution $f$.  This is not well known in
ICF plasmas, and is an active area of research.  For instance,
nonlocal transport due to scale lengths that are not sufficiently short compared
to collisional mean free paths, as well as hot electron generation by
SRS-produced LWs, lead to significant non-Maxwellian features. This becomes
more important for speeds larger than the thermal speed, where collisions become
less effective and which LW phase velocities generically are. The dominant
effect of non-thermal $f$ on our analysis is via the collisionless part of $\mathrm{Im}\chi$ and
the LW Landau damping rate,
which depends sensitively on $f(v_p)$, while $\mathrm{Re}\chi$
and the real frequency are determined by the bulk motion of the entire
$f$. The low $T_e$, high $n_e$ parameter region, with small $k\lde$ and large
$v_p/\vte$, is where the Landau damping is most susceptible to non-thermal
$f$. But the Landau damping is quite small here for a Maxwellian, and is
dominated by collisional damping.  The latter relies on the scattering of the bulk
electrons on ions, and is therefore not very sensitive to details of $f$. Our
results should be somewhat insensitive to the presence of non-Maxwellian tails.

We now consider the specific plasma conditions mentioned above as typical
for SRBS on NIF ignition experiments, namely $n_e/n_{cr}=0.1$ and $T_e=2$ keV
\cite{strozzi-srs-dpp-2011}. The phase-matched SRBS modes have $\la_1=553$ nm,
$k_2\la_{De}=0.297$, and $\om_2/\omp=1.155$. The calculated backscatter gain rate is significant in both
the CH ablator and He gas fill. The material affects a trapping assessment only
via collisions. From Fig.\ \ref{fig:dNrat}, for $\zeff=4$ we find $\delta N_c /
\delta N_{sl}=0.24$, so we just consider side loss. The 3D side loss detrapping
rate is $\nudsl=13.9$/ps, or a time of $1/\nudsl=0.072$ ps. The side loss threshold is
$\dNsl=2.7\times10^{-3}$, the critical intensity is $I_{cr}=1.96\times 10^{16}$
W/cm$^2$, and the threshold reflectivity is $\Rthrsl=(5.28\times 10^{13}$ W
cm$^{-2}/I_0)^2$. A typical intensity for inner cones of lasers in NIF ignition
experiments of $I_0=3\times10^{14}$ W/cm$^2$ gives $\Rthrsl=0.03$. Larger beam-averaged
reflectivities are frequently measured in experiments, and even larger values
will occur in intense speckles.

Finally, we show that smoothing by spectral dispersion (SSD) \cite{skupsky-ssd-jap-1989} is not likely to
reduce trapping effects in SRBS on NIF. Recent experiments have utilized $\De f_1=45$ GHz
of SSD bandwidth in the fundamental, 1054 nm laser light. After
frequency-tripling, this corresponds to a speckle lifetime of $t_{ssd}=1/(3\De f_1)=7.4$
ps. For the reference SRBS conditions discussed above, $t_{ssd}\sim100/\nudsl$. Thus SSD is
much less effective at detrapping than side loss.  Moreover, a Langmuir wave
overcomes SSD detrapping $(\tau_B<t_{ssd})$ for a very low amplitude of $\de N=2.5\times10^{-7}$, or
a reflectivity of $(4.9\times10^9$ W cm$^{-2}/I_0)^2$.

\section{Conclusions and future prospects} \label{s:conc}

This paper presented a framework for estimating when electron trapping
nonlinearity becomes important in Langmuir-wave dynamics. Detrapping by convective loss in the
longitudinal and transverse directions were discussed, as well as detrapping by
Coulomb collisions (electron-electron and electron-ion). 2D-2V simulations with
the Vlasov code \loki\ quantified trapping effects in
driven LWs with finite transverse profiles, and showed they increase with the
side loss bounce number as the transverse width increases.  These runs showed
trapping has little importance for $k\lde=0.7$, which is above the loss of
resonance value of 0.53. We explained this in terms of Eq.\ (\ref{eq:ET}), and related it
to the lack of kinetic inflation in non-resonant Raman scattering.

We applied our results to LWs driven by SRBS in NIF-relevant conditions. An
$f/8$ intense laser speckle was taken as the LW size in order to estimate side
loss. Side loss from speckles is generally more effective at detrapping
electrons than collisions, although this is not the case for wider
LWs or high-$Z$ plasmas. Linear response at the locally resonant SRBS scattered wavelength allowed
us to obtain an local reflectivity needed for trapping to overcome side loss.
This gives small values $(\sim1\%)$ for plasma conditions from which SRBS is thought
to originate in current NIF experiments.  Moreover, it is the speckle, and not
the lower beam average, intensity that matters.  Preliminary assessment of \ftd\
simulations \cite{hinkel-aps10-pop-2011} of NIF targets indicates a significant fraction of SRBS-generated
LWs is above our trapping threshold. Future work will assess this, and attempt to
incorporate trapping effects into enveloped propagation codes like \ftd.

\begin{acknowledgments}
  We gratefully recognize J.\ A.\ F.\ Hittinger and  R.\ L.\ Berger for
  helpful discussions and support. Work at LLNL was supported by US
  Dept.\ of Energy Contract DE-AC52-07NA27344 and partly under LDRD tracking
  number 12-ERD-061.
\end{acknowledgments}

\appendix
\section{Derivation of collisional results}

We restate our collision operator from Eq.\ (\ref{eq:fp}):
\begin{eqnarray}
  {1 \over \nu_0}{\p f \over \p t} &=& (1+\zeff)u^{-3}\p_\mu\lb(1-\mu^2)\p_\mu
  f\rb \nonumber \\
  &&+ 2u^{-2}\p_u(f+u^{-1}\p_uf) .
\end{eqnarray}
The second term describes collisions of tail electrons off bulk electrons, and
is valid for $u_p\gg1$. The parallel ($u_x$) and perpendicular ($\upe$)
velocities are given by $u_x=\mu u$ and $\upe =[1-\mu^2]^{1/2}u$ with
$u\in[0,\infty]$ and $\mu\in[-1,1]$. We compute $\ntr$, the number of electrons initially trapped,
that remain so up to time $t$. That is, once an electron leaves the trapping
region its coherent bounce motion stops, even if it re-enters the trapping
region later. The trapping region extends from $u_x=u_p\pm
u_{tr}/2$ and over all $\upe$.

Changing variables from $(\mu,u)$ to $(\upe ,u_x)$ gives 
\begin{eqnarray}
  {1 \over \nu_0}{\p f \over \p t} = \big[ D_{xx}\p^2_{u_x^2} + D_x \p_{u_x} + D_{x\perp} \p^2_{u_x\upe}
   \nonumber \\ + D_\perp \p_{\upe} + D_{\perp\perp}\p^2_{\upe^2} \big] f.
\end{eqnarray}
The $D$'s are straightforward to work out, and we do not give them. We assume
$u_{tr}$ is small, and order derivatives as $\p/\p u_x \sim 1/u_{tr}$ and $\p/\p
\upe \sim 1$ ($v_\perp \sim \vte$ in physical units). For sufficiently small
$u_{tr}$, the dominant term is $D_{xx}\p^2f/\p u_x^2$. For $u_p\gg1$, this is
valid if $u_{tr} \ll F(\zeff)/u_p$ where $F$ is a function of $\zeff$. With this
approximation, the collision operator yields a 1D diffusion equation:
\begin{eqnarray}
  \label{eq:18}
  {1 \over \nu_0}{\p f \over \p t} &=& D_{xx}{\p^2f \over \p u_x^2}, \\
 D_{xx} &\equiv& {(\upe^2+u_p^2)\upe^2(1+\zeff)+2u_p^2 \over (\upe^2+u_p^2)^{5/2}}.
\end{eqnarray}

We solve this equation subject to the outflow boundary conditions $f(u_x=u_\pm,\upe,t)=0$
with $u_\pm=u_p\pm u_{tr}/2$ the boundaries of the trapping region. The initial
condition for the trapped distribution is $f=f_0\exp[-(\upe^2+u_p^2)/2]$ (a Maxwellian with $u_x$ evaluated
at $u_p$) inside the trapping region, and $f=0$ otherwise. The number of trapped
electrons is
\begin{equation}
  \label{eq:7}
  \ntr = 2\pi \int_0^\infty d\upe\, \upe \int_{u_-}^{u_+} du_x f.
\end{equation}
We choose $f_0 = (2\pi u_{tr})^{-1} e^{u_p^2/2}$ so $N_{tr}(t=0)=1$. $f$ has the solution
\begin{eqnarray}
  \label{eq:8}
  f &=& \sum_{n=1,3,...} f_n(\upe,t)\sin n\pi w, \\
  f_n &=&  {2 \over \pi^2 n u_{tr}} \exp[-\upe^2/2 - n^2D\hat t].
\end{eqnarray}
$w \equiv (u_x-u_-)/u_{tr}$, $\hat t$ is given by Eq.\ (\ref{eq:thatcoll}), and
$D \equiv D_{xx}u_p^{-3}$. The sum is over odd positive integers since
the even terms vanish. The trapped fraction becomes
\begin{equation}
  \label{eq:9}
  \ntr =  {8 \over \pi^2} \sum_n n^{-2} \int_0^\infty dx\exp[-x-n^2D\hat t].
\end{equation}
$x \equiv \upe ^2/2$ is a  dummy integration variable. The decay rate of mode $n$ goes like $n^2$, as is typical
of diffusion problems.  After a short time, the $n=1$ term dominates. Retaining
just this term, and evaluating $8/\pi^2=0.81$, we find
\begin{eqnarray}
  \ntr &\approx& 0.81 I, \\
  I &\equiv& \int_0^\infty dx\, e^{-W}, \\
  W &=& x+Dt, \\
  D  &=& { (1+Z)2x(1+2ax)+2 \over (1+2ax)^{5/2} }. \label{eq:D}
\end{eqnarray}
To alleviate notation, we replaced $\hat t$ with $t$, $\zeff$ with $Z$, and defined $a \equiv
u_p^{-2}$. 

The upshot is an implicit integral equation for $t$:
\begin{equation}
I(t,a) = b \equiv {\ntr \over 0.81}.
\end{equation}
Numerically finding $t$ reveals it is linear in $a$ for $a<1$. We thus
write $t=t_{0}+at_{1}$ and expand for $a\ll1$:
\begin{equation}
  I \approx I(t_0,0) + at_{1}\p_tI(t_0,0) + a\p_aI(t_0,0) + O(a^2) = b.
\end{equation}
We choose $t_{0}$ such that $I(t_{0},0)=b$. We find
\begin{eqnarray}
  I(t,0) &=& \int dx \exp[-W_0], \\
  W_0 &=& x+D_0t, \\
  D_0 &=& 2(1+Z)x+2.
\end{eqnarray}
Performing the integral gives an implicit equation for $t$:
\begin{equation}
  \label{eq:t0imp}
  {e^{-2t} \over 1+2(1+Z)t} = b.
\end{equation}
This gives an exact formula for $\ntr$ in the limit $u_p\rightarrow\infty$:
\begin{equation}
  \label{eq:Ntrc_upinf}
  \ntr(t) = {0.81e^{-2t} \over 1+2(1+Z)t} \qquad u_p\rightarrow\infty.
\end{equation}
The temporal decay of $\ntr$ is thus not strictly exponential. This formula
reflects the different mathematical character of parallel dynamics from
electron-electron collisions $\rightarrow e^{-2t}$ and pitch-angle scattering from
collisions with all species $\rightarrow (1+Z)t$. Eq.\ (\ref{eq:t0imp}) is
transcendental, and can be ``solved'' in terms of the Lambert $W$ function. We
are interested in cases where $t<1$, so we Taylor expand $e^{-2t}$ to order
$t^2$ and obtain
\begin{equation}
  \label{eq:t0quad}
  t_0 = {1-b \over Y+ \lb Y^2-2+2b \rb^{1/2}}, \qquad Y\equiv 1+b(1+Z).
\end{equation}
This formula is valid ($t_0$ real) for $b$ above $b_0(Z)$. For
$Z=0$ we have $b_0=5^{1/2}-2 \approx 0.236$, and $b_0$ decreases with
$Z$. For our choice of $\ntr=1/2$, $b=0.617>b_0$ for all $Z$. We have
used the quadratic formula in a form that demonstrates the large-$Z$ limit
more clearly, which to leading order in $Y$ is
\begin{equation}
  t_0 \approx {1-b\over 2Y}, \qquad Y\gg1.
\end{equation}
With $\ntr=1/2$ this becomes
\begin{equation}
  t_0 \approx {0.31 \over 2.62+Z}.
\end{equation}
This form is accurate to within 10\% for all $Z \geq 0$. The correction for
finite $a$ is
\begin{eqnarray}
  t_1 &=& \left. -[\p_aI / \p_tI] \right| _{t=t_0,a=0} \\
     &=& -t_0 {\int_0^\infty dx\, \exp[-W_0](\p_aD)\left. \right|_{a=0} \over \int_0^\infty dx\, \exp[-W_0]D_0 }.
\end{eqnarray}
The result is
\begin{equation}
  \label{eq:t1}
  t_1 = t_0 {11+6Z+10t_0(1+Z) \over (1+2t_0(1+Z))(2+Z+2t_0(1+Z))}.
\end{equation}
Using $\ntr=1/2$ and our approximate form for $t_0$,
\begin{equation}
  t_1 \approx {1.15 + 5.70Z^{-1} + 6.09Z^{-2} \over Z + 7.23 +
    16.3Z^{-1} + 11.7Z^{-2}}.
\end{equation}

\subsection{Validity of Fokker-Planck (FP) Model}
\begin{figure}
  \centering
  \includegraphics[width=3in]{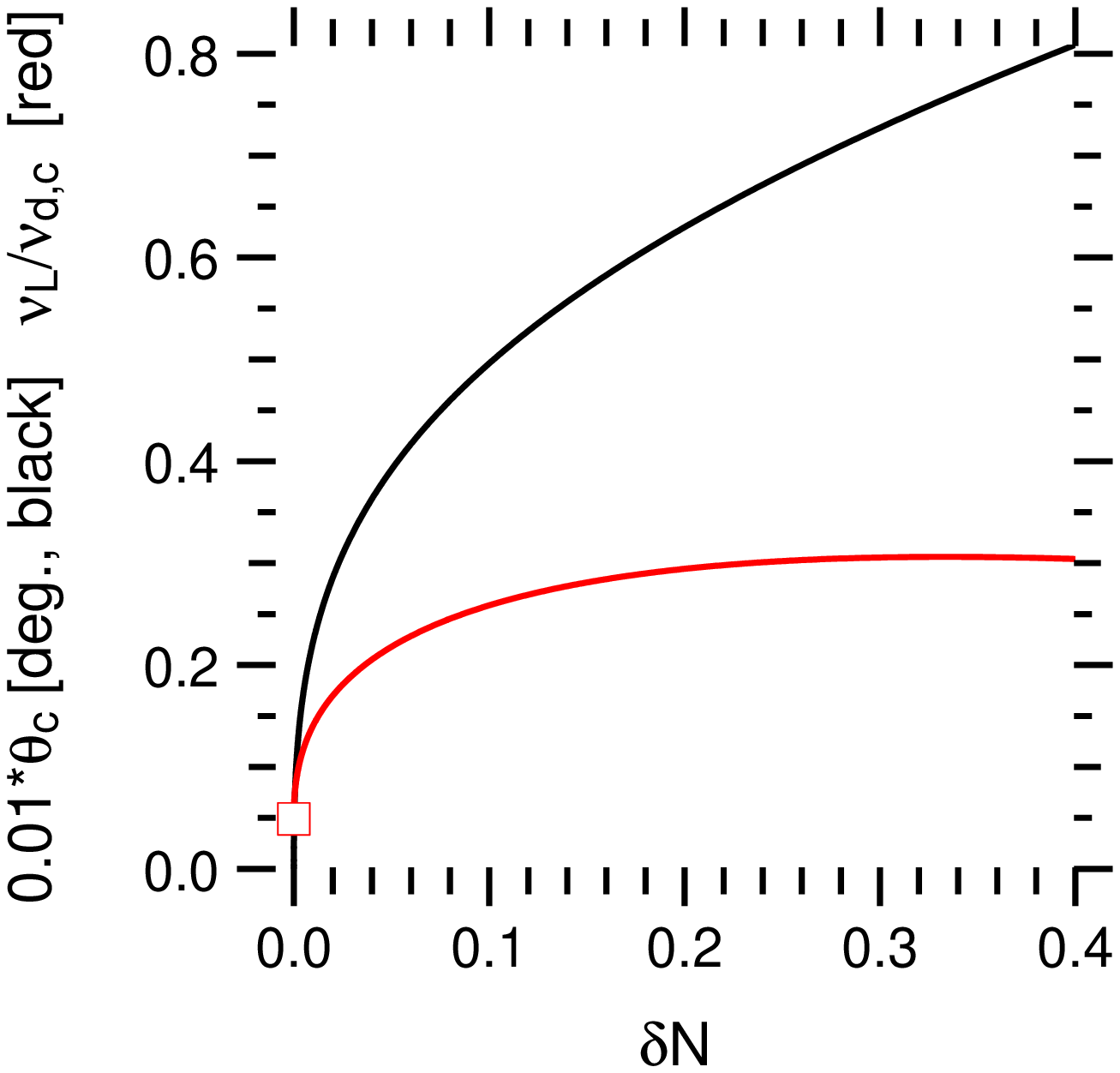}
  \caption{\colonl\ Black: angle $\ta_c$ separating small- from large-angle scattering from Eq.\
  (\ref{eq:tacr}). Red: Ratio $\nu_L/\nudc$ from Eq.\ (\ref{eq:nuLnudc}), with
  the $\de N=0$ value marked.}
  \label{fig:collval}
\end{figure}

Our FP model neglects large-angle scattering, which can detrap electrons in a
single collision. We estimate their importance, and show that the FP detrapping
rate dominates. As an example, we use the case from the end of Sec.\
\ref{s:icf}, namely $n_e/n_{cr}=0.1$, $T_e=2$ keV, $\la_1=$ 553 nm,
$k_2\lde=0.297$, $\om_2/\omp=1.155$, and $\zeff=4$, giving
$\ln\La_{ei}=7.5$. Consider a trapped electron with $v_x=v_p$ and a typical
$v_\perp=\vte$, which is elastically scattered ($|\vec v|=(v_p^2+\vte^2)^{1/2}=$
const.) to the boundary of the trapping region in one collision.  The electron's
(initial, final) angle with respect to the $\hat v_x$ direction is
$(\ta_I,\ta_F)$:
\begin{eqnarray}
  \label{eq:10}
  \cos\ta_I &=& {u_p \over \lp u_p^2+1\rp^{1/2} }, \\
  \cos\ta_F &=& {u_p-u_{tr}/2 \over \lp u_p^2+1\rp^{1/2} }. 
\end{eqnarray}
The critical angle $\ta_c=\ta_F-\ta_I$ separates large
from small scattering angles, and is given without approximation by
\begin{eqnarray}
  \label{eq:tacr}
  2\de N \cot^2{\ta_c\over 2} = -2\de N+ 2{\om\over\omp}\de
  N^{1/2}+(k\lde)^2 + \nonumber \\ k\lde \lb(k\lde)^2+4(\om/\omp)\de N^{1/2}-4\de N\rb^{1/2}.
\end{eqnarray}
Figure \ref{fig:collval} shows $\ta_c$ for our example parameters. For $\de N\ll1$, we have
\begin{equation}
  \label{eq:12}
  \ta_c \approx {2\de N^{1/2} \over \lb(k\lde)^2+2(\om/\omp)\de N^{1/2}\rb^{1/2}}.
\end{equation}

We employ the potential for Yukawa-screened Coulomb
scattering of an electron by an ion of charge $Z$:
$V=-(U_e/r)e^{-r/\lde}$ with $U_e\equiv Ze^2/4\pi\ep_0$. The quantum
cross-section \cite{griffiths-qm-1995}, in the first Born approximation, is
\begin{equation}
  {1\over\si_T}{d\si\over d\Om} = {1+\de\over4\pi} {\de \over (\de+\sin^2(\ta/2))^2}.
\end{equation}
$\si_T\equiv 4\pi(1+\de)^{-1}(U_e/\hbar\omp u_p)^2$ is the total cross-section,
and $\de\equiv(\hbar\omp/2u_pT_e)^2$ is unitless and typically small.  For our
example parameters, $\de=5.14\times10^{-9}$. The cross-section, integrated from $\ta_1$ to $\ta_2$, is
\begin{equation}
  {\si|^2_1 \over \si_T} = (1+\de)\de{\sin^2(\ta_2/2)-\sin^2(\ta_1/2) \over (\de+\sin^2(\ta_1/2))(\de+\sin^2(\ta_2/2))}.
\end{equation}
The total cross section from $\ta_1=0$ to $\ta_2=\pi$ is finite (without imposing any
cutoffs) and equals $\si_T$. The cross sections for small-angle
scattering $\si_S$ ($\ta_1=0$ to $\ta_2=\ta_c$) and large-angle scattering
$\si_L$ ($\ta_1=\ta_c$ to $\ta_2=\pi$) have the ratio
$\si_L/\si_S=\de(1+\de)^{-1}\cot^2(\ta_c/2)$. For our example parameters
 $\si_L<0.01\si_S$ for $\de N>5\times10^{-8}$. There are many more small- than
large-angle scatters, which is necessary for a FP model to be valid.

The detrapping rate due to large-angle scatters is approximately the rate at
which our typical electron undergoes one such scatter, i.e.\ $\nu_L = n_i \si_L v$.  We
compare $\nu_L$ for $1+\de\approx1$ to the FP detrapping rate, just due to electron-ion collisions and
using $\hat t \approx \hat t_0$:
\begin{eqnarray}
  \label{eq:nuLnudc}
  {\nu_L \over \nudc} &=& {0.36 \over\ln\La_{ei}}{\de N\cot^2(\ta_c/2) \over
    (k\lde)^2} \\
 &\approx& {0.36 \over\log\La_{ei}} \qquad \de N\ll 1.
\end{eqnarray}
The ratio depends only on $\ln\La_{ei}$ for small $\de N$, so the FP result
captures the basic parameter dependence. Large-angle scattering enhances the FP
detrapping rate by a modest amount. For our parameters, $\nu_L/\nudc=0.048$ at $\de N=0$, and the ratio is plotted
in Fig.\ \ref{fig:collval}.  Since the FP results were found for $\de N\ll1$,
the comparison of $\nu_L$ and $\nudc$ may not be accurate at large $\de N$.



%

\end{document}